\documentclass[fleqn,usenatbib,useAMS,letterpaper]{mn2e_arxiv}
\usepackage{longtable}
\usepackage[applemac]{inputenc}
\usepackage[draft=false, colorlinks=true, allcolors=blue]{hyperref}
\usepackage{graphicx}	
\usepackage{amsmath}	
\usepackage{dsfont}
\usepackage{bm}
\usepackage{mathrsfs}


\usepackage[T1]{fontenc}
\usepackage{ae,aecompl}

\usepackage{newtxtext,newtxmath}

\title[High Precision Power Spectrum Emulation]{The Mira-Titan Universe IV. High Precision Power Spectrum Emulation}

\author[K. R. Moran et al.]{Kelly R. Moran\thanks{Contact e-mail:
    \href{mailto:krmoran@lanl.gov}{krmoran@lanl.gov}}$^{1}$,
  Katrin Heitmann$^{2}$,
  Earl Lawrence$^{1}$,
  Salman Habib$^{2,3}$,\newauthor
  Derek Bingham$^{4}$,
  Amol Upadhye$^{5}$, 
  Juliana Kwan$^{5}$,
  David Higdon$^{6}$, Richard~Payne$^{1}$
  \\ \\
  $^{1}$Statistical Sciences Group, CCS Division, Los Alamos National Laboratory, Los Alamos, NM 87545  \\
 $^{2}$High Energy Physics Division, Argonne National Laboratory, Lemont, IL 60439, USA\\
  $^{3}$Computational Science Division, Argonne National Laboratory, Lemont, IL 60439, USA\\
  $^4$Department of Statistics and Actuarial Science, Simon Fraser University, Bunraby, BC, Canada\\
  $^5$Astrophysics Research Institute, Liverpool John Moores University, Liverpool, United Kingdom\\
  $^6$Department of Statistics, Virginia Tech, Blacksburg, VA 24061\\}

\date{}

\pubyear{2022}

\begin{document}\label{firstpage}
\pagerange{\pageref{firstpage}--\pageref{lastpage}}
\maketitle

\begin{abstract}
  Modern cosmological surveys are delivering datasets characterized by unprecedented quality and statistical completeness; this trend is expected to continue into the future as new ground- and space-based surveys come online. In order to maximally extract cosmological information from these observations, matching theoretical predictions are needed. At low redshifts, the surveys probe the nonlinear regime of structure formation where cosmological simulations are the primary means of obtaining the required information. The computational cost of sufficiently resolved large-volume simulations makes it prohibitive to run very large ensembles. Nevertheless, precision emulators built on a tractable number of high-quality simulations can be used to build very fast prediction schemes to enable a variety of cosmological inference studies. We have recently introduced the Mira-Titan Universe simulation suite designed to construct emulators for a range of cosmological probes. The suite covers the standard six cosmological parameters $\{\omega_m,\omega_b, \sigma_8, h, n_s, w_0\}$ and, in addition, includes massive neutrinos and a dynamical dark energy equation of state, \{$\omega_{\nu}, w_a\}$. In this paper we present the final emulator for the matter power spectrum based on 111 cosmological simulations, each covering a (2.1Gpc)$^3$ volume and evolving 3200$^3$ particles. An additional set of 1776 lower-resolution simulations and TimeRG perturbation theory results for the power spectrum are used to cover scales straddling the linear to mildly nonlinear regimes. The emulator provides predictions at the two to three percent level of accuracy over a wide range of cosmological parameters and is publicly released as part of this paper.
\end{abstract}




\section{Introduction}
\label{sec:intro}

Over the last several decades, optical cosmological surveys have made key contributions to the revolution in our knowledge of the Universe and will continue to do so in the future. Observations carried out by, e.g., the Dark Energy Spectroscopic Instrument (DESI)~\citep{desi}, the Vera C. Rubin Observatory's Legacy Survey of Space and Time (LSST)~\citep{lsst,desc}, the Nancy Grace Roman Space Telescope ~\citep{wfirst}, the Euclid satellite~\citep{euclid} and the Spectro-Photometer for the History of the Universe, Epoch of Reionization and Ices Explorer (SPHEREx)~\citep{spherex} will continue the legacy of past surveys such as the CfA Redshift Survey\footnote{https://lweb.cfa.harvard.edu/~dfabricant/huchra/zcat/}, the Dark Energy Survey\footnote{https://www.darkenergysurvey.org/} (DES), 2dF Galaxy Redshift Survey\footnote{http://www.2dfgrs.net/}, KiDS\footnote{http://kids.strw.leidenuniv.nl/}, and the Sloan Digital Sky Survey\footnote{https://www.sdss.org/} (SDSS). A major focus of next-generation surveys is to further our understanding of the origin of late-time cosmic acceleration, which, at this point, remains fundamentally mysterious. The upcoming large-scale structure data sets will also lead to improved constraints (possibly measurements) on the neutrino mass sum and hierarchy options, exemplars of fundamental science targets for which cosmology can provide uniquely powerful probes.

Contemporary and future surveys aimed at unraveling some of the mysteries of the `Dark Universe' face a major challenge with regard to accessing accurate predictions for a number of observables. These predictions are critical to interpret survey data and to fully unlock the information that is potentially available on small scales. As a key example, the matter power spectrum is a fundamental ingredient in a number of cosmological analyses. Early papers (e.g. \citealt{2005APh....23..369H}) estimated the accuracy with which the power spectrum must be modeled to achieve percent level cosmological parameter estimates from surveys. It was immediately clear that simulations capturing the nonlinear regime of structure formation would be far too computationally expensive to be directly useful in solving the inverse problem of parameter estimation from observational data.

In order to address this challenge, fitting approaches were first developed; a well-known example is Halofit~\citep{smith03}, based on a functional form inspired by structure formation physics. A set of 30 fitting parameters were optimized using $\Lambda$CDM ($\Lambda$ Cold Dark Matter) simulations to estimate the nonlinear power spectrum. \cite{bird} added massive neutrino parameters to the fit. The Halofit~\citep{smith03} approach was also refined by \cite{takahashi12} using 16 new, higher-resolution simulations and 35 free parameters. This led to predictions accurate at the $\sim$6\% level close to currently best-fit $\Lambda$CDM models and $\sim$10\% further away for more general $w$CDM models. Another example is  HMCODE-2020, first described in \cite{mead16}, which also uses the halo model as a starting point; the most recent version \citep{2021MNRAS.502.1401M} allows for the inclusion of baryonic effects and improved predictions for neutrinos. While fitting functions are easy to use in analysis work, the (uncontrolled) complex form of the fit and the limited accuracy obtained in spite of the large number of fitting parameters used, both point to the difficulties inherent in this approach when confronting upcoming survey data.

More than a decade ago our group introduced the idea of constructing surrogate models or {\em emulators} based on a small set of high-quality simulations~\citep{HHHN,HHHNW}. A full end-to-end example of the approach combining emulation and combined cosmological parameter calibration from large scale structure (SDSS) and cosmic microwave background (CMB) data is described in \cite{higdon10}. The paradigm for emulator construction relies on an efficient sampling design that covers the cosmological parameter space of interest with only a (relatively) small number of models along with sophisticated interpolation schemes that provide predictions anywhere within the parameter space at high accuracy. The accuracy of the predictions depends in turn on the accuracy of the simulations and the coverage of the desired cosmological parameter space by the simulation suite. 

We used surrogate models in \cite{HHHN} and \cite{HHHNW} to build accurate emulators for matter and CMB power spectra. Following this proof of concept work, we carried out the Coyote Universe project~(\citealt{coyote1,coyote2,coyote3}); the papers established criteria for the accuracy of the simulation set-up~\citep{coyote1}, tested the accuracy of the emulator construction with surrogate models~\citep{coyote2}, and delivered the final, first high-accuracy power spectrum emulator~\citep{coyote3}. The emulator design covered the five parameters of the cosmological standard model and also allowed for a variation of the value of the dark energy equation of state parameter $w$. In~\cite{emu_ext} the work was extended to include one additional cosmological parameter and to expand the $k$-range 
covered by the predictions. The suite of Coyote Universe simulations was also used to provide emulators for the concentration-mass relation \citep{2013ApJ...768..123K} and the galaxy power spectrum~\citep{2015ApJ...810...35K}.

The construction of emulators, as an alternative to fitting functions, has now become an entrenched methodology and a number of new ideas have been introduced, along with new applications. The {\sc Aemulus} project~\citep{aemulus1,aemulus2,aemulus3,aemulus4} carried out a set of simulations spanning a 7 dimensional parameter-space to build a range of emulators, including predictions for the mass function, halo bias, and the galaxy correlation function. The Euclid Collaboration has embraced the emulator concept to prepare for their cosmology analysis and has focused so far on predictions for the power spectrum~\citep{2021MNRAS.505.2840E,2019MNRAS.484.5509E}. In the BACCO project~\citep{2021MNRAS.507.5869A}, emulation is used to provide predictions for the matter power spectrum with a small number of simulations augmented by a rescaling approach to generate predictions for a larger set of cosmologies. In follow-up work, baryonic effects have been added to the emulator~\citep{2021MNRAS.506.4070A}, and a new emulator for the two-point clustering measurements of biased tracers has been released~\citep{2021arXiv210112187Z}. In \cite{2021arXiv211102419M}, an emulator approach was employed to derive cosmological constraints from weak lensing and galaxy clustering measurements from Hyper Suprime-Cam Year-1 (HSC-Y1) data and SDSS. This work is based on the Dark Emulator package, presented in~\cite{2019ApJ...884...29N}. The Dark Emulator provides predictions for a range of cosmological statistics.

With the advent of more capable supercomputers and encouraged by the success of the Coyote Universe project, we designed a more extensive simulation campaign to increase the cosmological parameter space covered by the emulators and to considerably improve the quality of the simulations with regard to volume and resolution. The new campaign, named the Mira-Titan Universe after the supercomputers the simulations were carried out on (both since retired), was first described in~\cite{heitmann15}. A new feature of the Mira-Titan Universe simulations was a nested design strategy that allows for sequentially adding simulations that systematically improve the accuracy of the resulting emulators, something not possible with conventional sampling strategies such as Latin hypercube sampling. The first power spectrum emulator based on a subset of 36 cosmological models is described in~\cite{MT-pow1}. Later, a mass function emulator based on the full simulation suite covering 111 cosmological models was released~\citep{MT-massf}. We also investigated the (cluster-scale) aperture mass function in \cite{2022arXiv220316379D}.

In this paper, we follow a strategy very similar to that of~\cite{MT-pow1} to build the final power spectrum emulator based on the full simulation suite. It is interesting to note that controlled accuracies at the per cent level can be obtained with a sampling sparsity of less than two points per parameter space dimension, {\em without the use of any additional modeling assumptions}. Moreover, because of our sampling method, further improving emulation accuracy over a subspace of the parameter hypercube is possible by adding more perturbation theory and simulation results in a principled manner, such that the eventual accuracy is limited only by the ultimate fidelity of the theoretical and simulation inputs.

It is important to emphasize the nature and volume of the considered parameter-sample space. Our aim is to achieve an emulation strategy that is largely unbiased over the sampled parameter hypercube, with errors becoming (unavoidably) larger near the hypercube edges. The idea is to have a more or less flat prior over the considered volume of parameter space, with the assumption that, when the emulator is used in an inverse problem, the estimated parameter values will lie reasonably within the interior of the hypercube. Clearly, restricting the parameter hypercube volume makes it easier to build emulators (fewer samples, higher accuracy), but also runs the risk of increased bias and overall inaccuracy in parameter estimation applications, if one strays closer to, or even goes beyond, sufficiently well-sampled parameter space boundaries. For this reason, our parameter range has been kept deliberately broad. As one illustrative example, our wide range in $h$ ($0.55-0.85$) comfortably includes the current `Hubble tension' range (see, e.g., Figure~10 in \citealt{Freedman2021}), which a more aggressive choice would exclude (e.g., $0.61 \leq h \leq 0.73$ in the emulator described in \citealt{2021MNRAS.505.2840E}). Similarly, techniques based on incorporating known results from Markov Chain Monte Carlo (MCMC) chains to reduce the volume within the hypercube (e.g., \citealt{aemulus1}) may not be useful in applications where samples outside the considered limited region are needed; this can easily happen when using different probes (or a smaller number thereof) to constrain cosmological parameters with respect to the original choices that led to a particular restriction of the hypervolume.

The matter power spectrum emulator released with this paper covers a redshift range of $z=0$ to $z=2.02$ with a maximum wave number of $k=5$~Mpc$^{-1}$. We note that the predictions are from gravity-only simulations and baryonic effects (e.g., gas cooling, astrophysical feedback mechanisms) are expected to affect the power spectrum in a nontrivial way at $k$ values of greater than unity, at the accuracies considered in this paper. Unfortunately, these effects are uncertain and difficult to model accurately. One value of a precision prediction from a gravity-only approach is that it can be compared to (sufficiently small scale) observations and provide a direct estimate of the size and possible functional shape of the baryonic modification, which can then be studied separately with hydrodynamical simulations, or modeled in other ways.

The rest of the paper is organized as follows. In Section~\ref{sec:sims} we provide an overview of the Mira-Titan simulation suite and Time-RG perturbation theory results underlying this work and detail the generation of the power spectra from the simulation suite, including a description of how the different results are combined (perturbation theory and lower- and higher-resolution simulations). Next, in Section~\ref{sec:emu-construct}, we describe the emulator construction, essentially following the procedure described in~\cite{MT-pow1}, with some minor variations. In Section~\ref{sec:emperf} we present and discuss different types of results for assessing emulator performance. We evaluate performance via a set of independent simulations, internal out-of-sample tests, and comparisons with some other independent power spectrum prediction methods based on fitting functions as well as emulator-based approaches. We summarize our overall approach and findings in Section~\ref{sec:sum} and provide an outlook for future work. All model parameters are specified in the Appendix.

\section{The Mira-Titan Universe Simulation Suite}
\label{sec:sims}

The Mira-Titan Universe simulation suite was first introduced in~\cite{heitmann15}. In this section we briefly summarize the major features and simulations that are particularly relevant for the generation of the matter power spectrum emulator, along with relevant changes in data processing and management since the work reported in~\cite{MT-pow1}. A more comprehensive description of the implementation details of, e.g., massive neutrinos is available in~\cite{MT-pow1}. All simulations were carried out with HACC, the Hardware/Hybrid Accelerated Cosmology Code~\citep{habib16, habib17} running in its CPU (Mira) and CPU/GPU modes (Titan) with separately optimized force computation algorithms in the two cases. 

\subsection{Suite Design}

The simulation suite covers eight cosmological parameters over the following ranges:
\begin{subequations}
\label{eqn:paramranges}
\begin{eqnarray}
0.12\le &\omega_m& \le 0.155,\\
0.0215\le &\omega_b& \le 0.0235,\\
0.7\le &\sigma_8& \le 0.9,\\
0.55\le &h& \le 0.85,\\
0.85\le &n_s& \le 1.05,\\
-1.3\le &w_0& \le -0.7,\\
-1.73\le &w_a& \le 1.28,\\
0.0\le &\omega_\nu& \le 0.01,
\end{eqnarray}
\end{subequations}
with a joint constraint on $w_0$ and $w_a$ to obey $w_0+w_a\leq-0.0081$. This constraint was imposed to avoid unphysical realizations for the dark energy equation of state. The effective number of neutrino species is fixed to be $N_{\rm eff}=3.04$ and we restrict our investigations to flat geometries, $\Omega_k=0$ (`inflation prior'). For more details on the parameter choices and their ranges, see \cite{heitmann15}. Outputs at eight different redshifts $z\in\{0.00, 0.10, 0.24, 0.43, 0.66, 1.01, 1.61, 2.02\}$ were used in the emulator construction to provide a redshift range of $0\leq z \leq 2.02$.

As discussed in detail in \cite{heitmann15}, the simulation design (the cosmological models to be simulated) is based on a nested, space-filling lattice. The advantage behind the approach is that a first set of models can provide sufficient coverage of the parameter space to enable the construction of a reasonably faithful emulator. More models can then be added sequentially to the original set to improve the accuracy further (this is not possible, for example, with Latin hypercube sampling, where an entirely new set of simulations would be needed, see, e.g.,~\citealt{HHHNW}). In our specific case, we first chose 26 models for the base design. These models were augmented with an additional 10 models having massless neutrinos, following a separate symmetric Latin hypercube design in the same parameter range as given above. The additional 10 models were included to enable adequate coverage for models that obey the current `Standard Model' of cosmology, which assumes $\Omega_\nu=0$. In \cite{MT-pow1}, these first 36 models were used to build an emulator for the matter power spectrum that is accurate at the 6\% level. The second design stage added 29 additional models, and finally the full model space used in this paper covers 111 models across eight redshifts $z\in\{0.0, 0.101, 0.242, 0.434, 0.656, 1.006, 1.61, 2.02\}$. All models are listed in the Appendix. \cite{MT-massf} provide a visual impression of the model space covered in their Figure~1. 

\subsection{Time-Renormalization Group Perturbation Theory}

At low values of the wave number, $k$, corresponding to large length scales, the Mira-Titan emulator relies upon the Time-Renormalization Group (Time-RG) perturbation theory of~\cite{Pietroni:2008jx} and \cite{Lesgourgues:2009am}.  Briefly, fluid dynamics in an expanding universe in general relativity is well-approximated deep inside the horizon by classical fluid dynamics in an expanding volume.  The continuity and Euler equations of classical fluid dynamics, applied to $N$-point correlation functions, imply an infinite hierarchy of evolution equations relating the time-derivative of each $N$-point correlator to $(N+1)$-point correlators.  Time-RG truncates this hierarchy after $N=3$ and then integrates these equations for each wave number, thereby making the bispectrum the repository of nonlinear clustering information.  If massive neutrinos are present, Time-RG treats them linearly using a precomputed neutrino-to-total-matter clustering ratio~\citep{Upadhye:2013ndm}.

\subsection{Handling Neutrinos}

Due to its large thermal velocity dispersion, the neutrino population has a characteristic length scale, the free-streaming scale, dividing its perturbative behavior into two distinct regimes. At distances much larger than the free-streaming length, neutrino thermal motion becomes negligible, and the neutrino density contrast approaches that of the total matter, $\delta_\nu(k) / \delta_\mathrm{m}(k) \rightarrow 1$.  Well below the free-streaming length, linear neutrino clustering is suppressed as $\delta_\nu(k) / \delta_\mathrm{m}(k) \rightarrow k_\mathrm{FS}^2 / k^2$, with the free-streaming wave number $k_\mathrm{FS} = \sqrt{4\pi G a^2 \bar\rho_\mathrm{m}} / c_\nu$ and $c_\nu^2 = 3 \zeta(3) T_{\nu,0}^2 / (2 \ln(2) m_\nu^2 a^2)$  as in \cite{Ringwald:2004np}. This small-scale growth suppression is imprinted upon all other clustering species, making the linear growth factor of the CDM+baryon (cb) fluid scale-dependent.

The Mira-Titan simulations neglect the feedback of neutrino clustering on CDM~+~baryon clustering.  In effect, they treat the neutrinos as free streaming at all scales, an approximation justified by the fact that most cosmological information is in the neutrino free-streaming regime.  Consistently joining the simulated power spectra to the perturbative ones at large scales requires that we correct for this approximation by multiplying each simulated power spectrum by the squared growth factor ratio $D_\mathrm{cb}(k)^2 / D_\mathrm{cb}(k_\mathrm{high})^2$, where $k_\mathrm{high}$ is chosen to be much larger than $k_\mathrm{FS}$.  Since neutrino clustering above the free-streaming length enhances clustering of the CDM~+~baryon fluid, this correction factor is larger than one at low $k$ and scale factor $a$ while approaching unity as $k$ exceeds $k_\mathrm{FS}$, as shown in~\cite{Upadhye:2013ndm}.  The current release of the Mira-Titan emulator improves the accuracy of this scale-dependent growth correction and enlarges the scale at which simulated power spectra are aligned to perturbative ones.

\subsection{Data Alignment}
Before the various power spectrum measurements are combined, we interpolate the perturbation theory (PT), lower-resolution (LR), and higher-resolution (HR) runs onto a common set of wavenumbers using a cubic spline. This is necessary because the $k$-spacing for the N-body power spectra are determined by the size of the simulation and the Fast Fourier Transform (FFT) grid, which differs between the LR and HR runs. (Part of the data alignment procedure -- the LR to HR runs -- is also discussed in \citealt{emu_ext}.)

\begin{figure}
\centering
\includegraphics[width=\columnwidth]{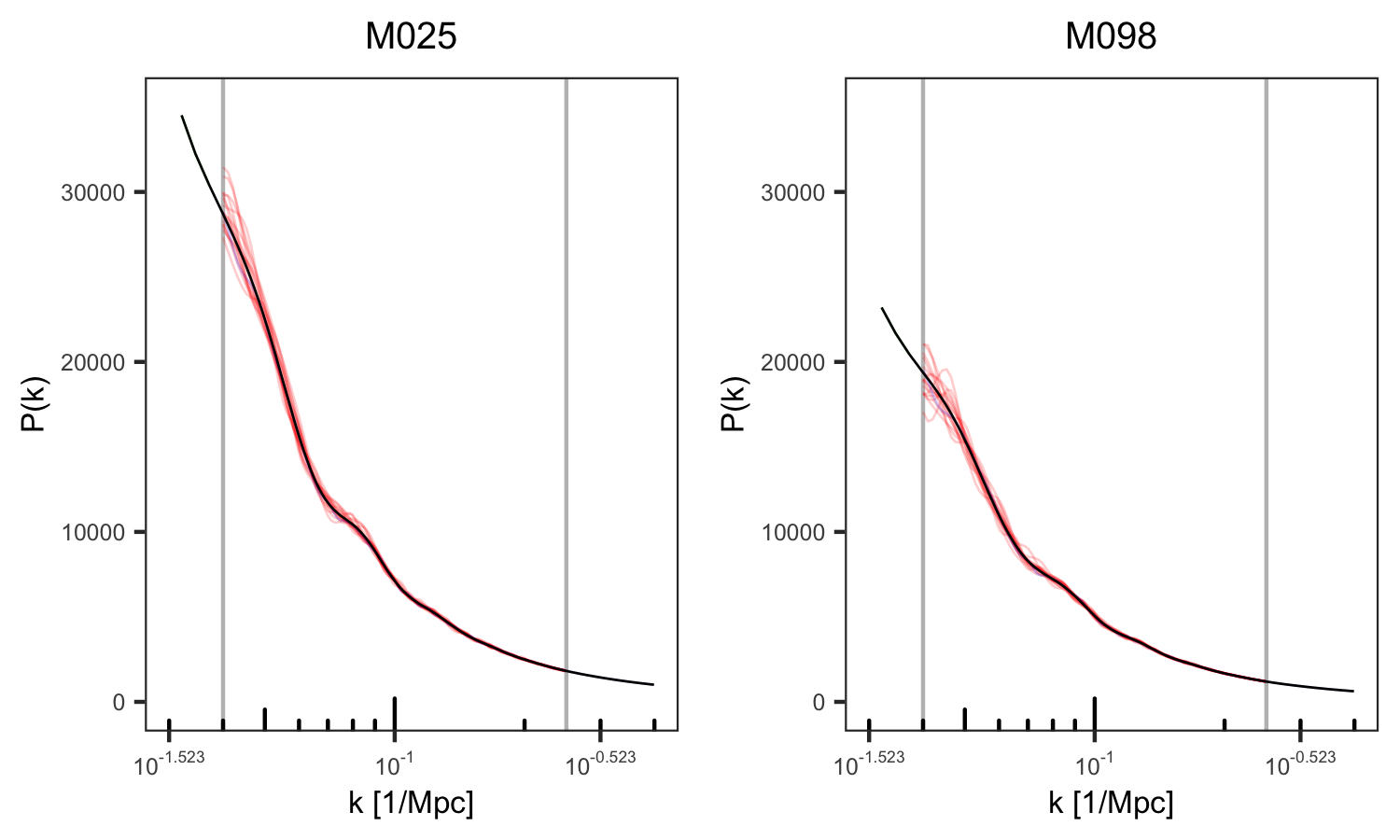}
\caption{Simulated power spectra and their associated smooth estimates for two cosmologies at an illustrative redshift ($z=0.656$). The dashed grey lines indicate the point at which we stop using perturbation theory and begin using simulated spectra (left, $k=0.04$ Mpc$^{-1}$), and the point at which we stop using the lower-resolution simulations (right, $k=0.25$ Mpc$^{-1}$).}
\label{fig:data_smooth_spec}
\end{figure}

We then perform two data alignment steps across the relevant $k$-windows prior to emulator construction, new to the Mira-Titan suite (since~\citealt{MT-pow1}). The first alignment primarily adjusts for the finite size of the LR simulation boxes, which while large (1.3~Gpc to a side), still have an associated $P(k)$ finite-size suppression of about $\sim 0.1$\% at the smallest $k$ values. The second step aligns the HR spectra to the LR spectra in the same way as in \cite{emu_ext}. These are small adjustments, significantly smaller than the per cent level error for which the emulator is designed. In carrying out the adjustments, we use the average $\text{P}(k)$ over a (small) $k$-window to mitigate the issue of shifting the spectra too far just because of the direction of noise in the data at a particular point. Full alignment details are provided in the Appendix, and examples of spectra near the match windows before and after alignment are provided in the supplemental materials.

\begin{figure}
\centering
\includegraphics[width=\columnwidth]{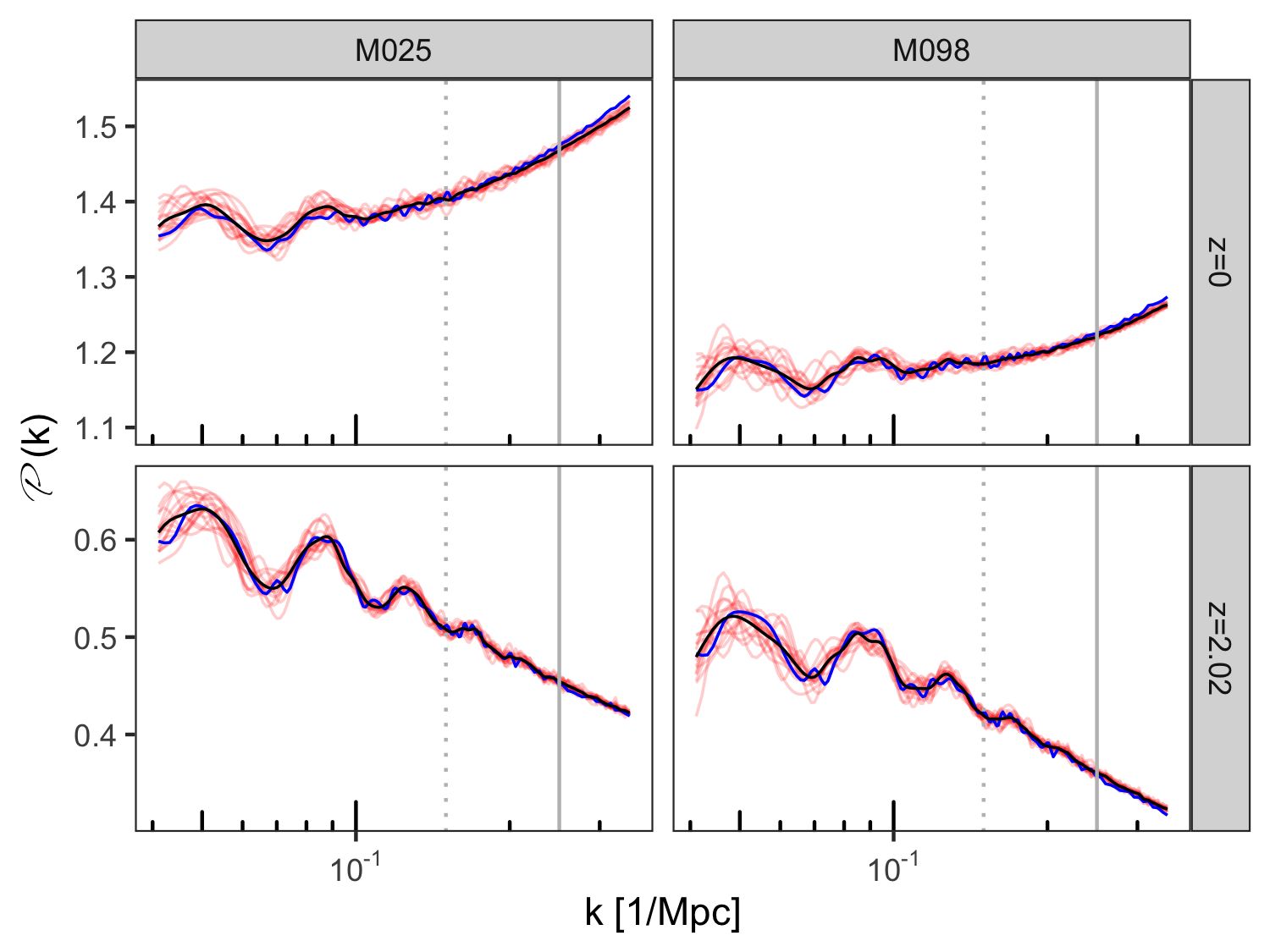}
\caption{Simulated power spectra on the emulation (transformed) space (Eqn.~\ref{trans}) for two example cosmologies at the minimum and maximum redshifts. LR simulations are shown as red lines, the average of these LR simulations is shown as a black line, and each HR simulation run (one per cosmology/redshift combination) is shown as a blue line. Subtle bumps in the power spectra can be seen between $k=0.15$~Mpc$^{-1}$ (dashed grey vertical line) and $k=0.25$~Mpc$^{-1}$ (solid grey vertical line). The power spectrum on the emulation (transformed) space is calculated as $\mathcal{P}(k) = \log_{10}(\Delta^2(k) / k^{1.5})$, where $\mathcal{P}(k)$ denotes the transformed spectrum.}
\label{fig:data_processing_bumps}
\end{figure}

\begin{figure}
\centering
\includegraphics[width=\columnwidth]{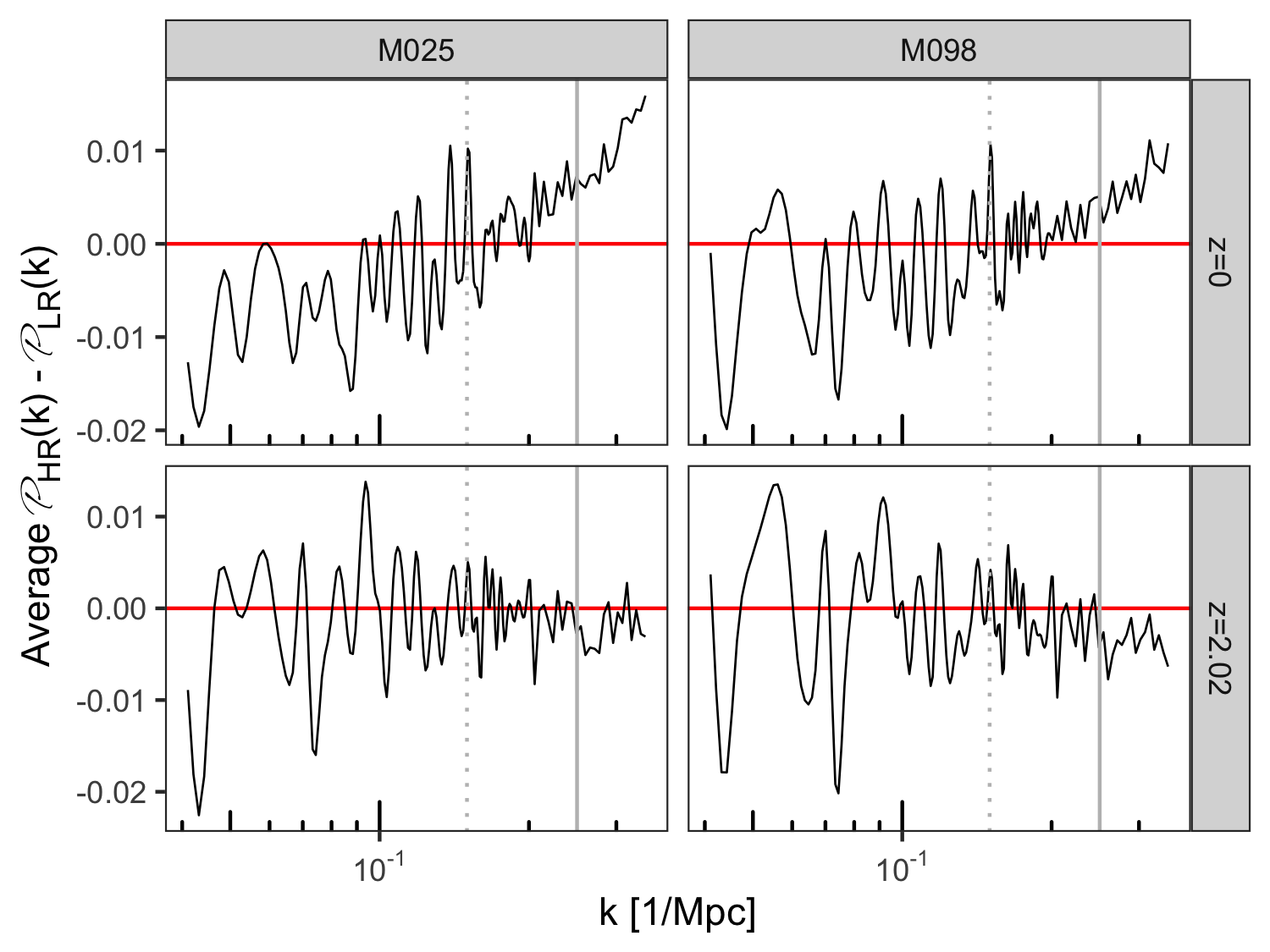}
\caption{Difference between the high-resolution (HR) and the average low-resolution (LR) power spectra in the emulation (transformed) space (Eqn.~\ref{trans}) for two example cosmologies at the minimum and maximum redshifts ($z=0$ and $z=2.02$). A systematic bias in the LR simulations can be discerned with increasing $k$, beginning to appear at around $k=0.15$~Mpc$^{-1}$ (dashed grey vertical line) for some cosmology/redshift combinations (e.g., M025 with $z=0$) but remains modest at $k$ below 0.25~Mpc$^{-1}$ (solid grey vertical line).}
\label{fig:data_processing_bumps_diff}
\end{figure}

The $k$-ranges at which each of the PT, LR simulations, and HR simulations are included are chosen to balance the validity of realizations (and possible bias introduction) with the ability to capture subtle signals in the power spectra (e.g., baryonic acoustic oscillations).
Cutoffs, simulation results, and smoothed spectra are visualized in Figure~\ref{fig:data_smooth_spec}. For $k<0.04$~Mpc$^{-1}$, we use only the PT output. Beyond $k=0.04$~Mpc$^{-1}$ we stop using the PT output and begin using both the LR and HR spectra. At $k=0.25$~Mpc$^{-1}$, we stop using the LR spectra, so that for $k \geq 0.25$~Mpc$^{-1}$ we only use the HR spectra. Figures~\ref{fig:data_processing_bumps} and \ref{fig:data_processing_bumps_diff} illustrate the trade-off between bias and signal capture of this choice. Figure~\ref{fig:data_processing_bumps} shows the spectra on the emulation (transformed) space for two example cosmologies at two illustrative redshifts, in which there are prominent bumps easily visible in the average of the LR simulations below the match point $k=0.25$~Mpc$^{-1}$ that are less clear in each singly-run HR simulation. Including the LR simulations below $k=0.25$~Mpc$^{-1}$ provides more data by which to estimate this underlying smooth signal, whereas only including each HR simulation after some lower $k$ (e.g., $k=0.15$~Mpc$^{-1}$, as shown in the plots) increases the likelihood that the noise overwhelms the signal in the region of interest. Figure \ref{fig:data_processing_bumps_diff} more clearly shows the downside of this higher cutoff -- there is systematic bias in the LR simulations with increasing $k$, which begins to appear at around $k=0.15$~Mpc$^{-1}$. However, this bias remains modest at $k$ values below $0.25$~Mpc$^{-1}$.

\section{Emulator Construction}
\label{sec:emu-construct}

The error present in predictions of the matter power spectrum is comprised of irreducible and reducible components. Here, the former is due to the numerical accuracy of the underlying cosmological simulations, which is at the percent level at the redshifts and scales of interest. The reducible error in our most recent emulator release~\citep{MT-pow1} was largely due to the limited number of models that were considered. As we will show, this error has been significantly lowered now that we have a complete set of runs from the full design.

Because we are working with the full design, the emulator construction is relatively straightforward compared to that of \cite{MT-pow1}, and nearly identical to the process described in \cite{coyote3}. We review that approach here. In brief, a {\em process convolution} model is used to combine the noisy output of the simulations for a given cosmology into a prediction of the smooth power spectrum for that cosmology. The smooth spectra are used to construct an {\em empirical orthogonal basis}, and {\em Gaussian process regression} is used to predict the weights for each basis as a function of the cosmological parameters.

As in \cite{MT-pow1}, we transform each of the simulated spectra prior to the process convolution step in order to accentuate the baryonic acoustic oscillations.
Specifically, 
\begin{equation}
    \mathcal{P}(k) = \log_{10}\big({k^{1.5} P(k)}/{2\pi^2}\big) = \log_{10}(\Delta^2(k) / k^{1.5}),
    \label{trans}
\end{equation}
where $\mathcal{P}(k)$ denotes the transformed spectrum. After emulation, predictions are transformed back into the original space. Results shown in the following sections are in the original space unless otherwise noted.

\subsection{Estimating the Smooth Power Spectra}
\label{sec:smooth}

Prior to emulation, we use a two-layer process convolution model to estimate the unknown smooth power spectra for each cosmology and redshift based on the noisy simulated spectra.  A process convolution (\citealt{higdon02}) is a continuous stochastic process generated by convolving a simple, possibly discrete, process with a kernel -- see Figure 2.1 from \cite{higdon02} for an example. In a two-layer process convolution, the kernel width parameter is also modeled using a process convolution. Such an approach is useful because it allows for multiple observations of a function, that may cover different regions and have different amounts of noise, to be combined to generate a smooth estimate. The second layer of the process enables the ``wiggliness" of the function to vary across values of $k$. This approach is the same as that taken in our previous work, including \cite{coyote2} and \cite{MT-pow1}, and specification of the process, with notation shared with that used in \cite{coyote2}, is described below.

\begin{figure}
\includegraphics[width=\columnwidth]{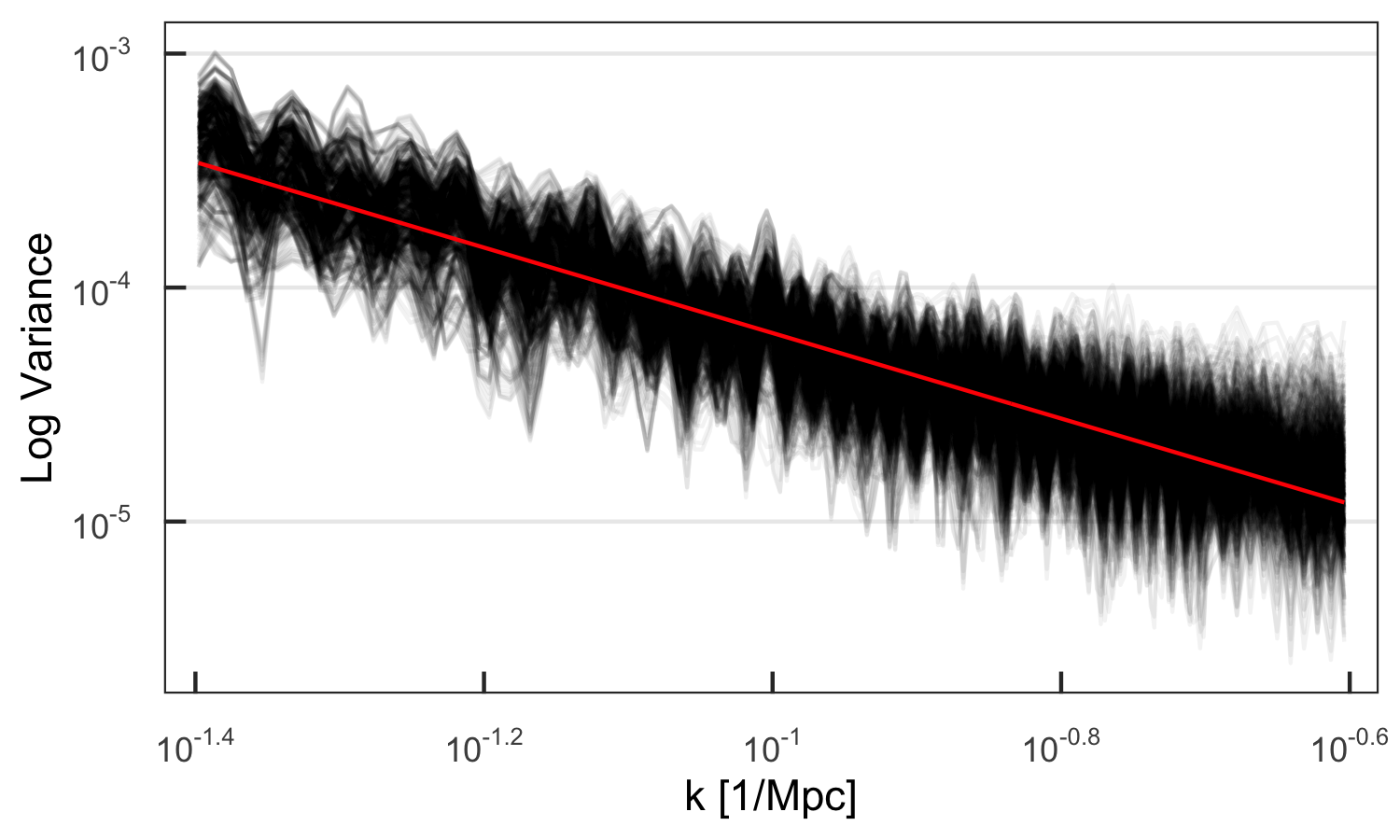}
\includegraphics[width=\columnwidth]{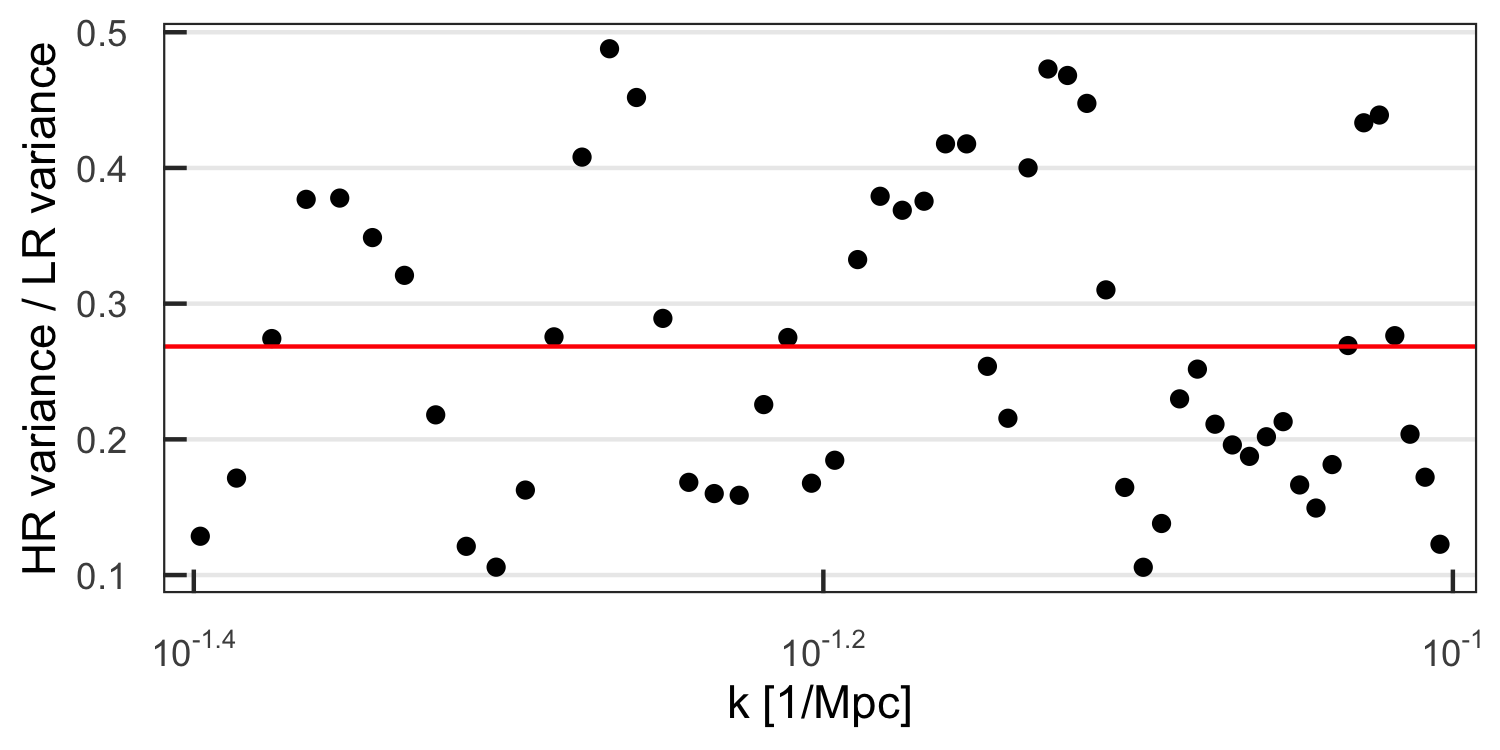}
\caption{Upper panel: Each black line shows the logged LR variance for a single cosmology/redshift combination, calculated at each $k$ by using the replicate LR simulations. The log-log regression modeled LR variance by $k$ is shown as the straight red line. Lower panel: Each point shows the ratio of the HR variance to the LR variance calculated at a single $k$ by using all cosmology/redshift combinations. The HR variance is calculated using the average of the LR simulations as the mean at each cosmology and redshift. The average of this ratio is 0.268, shown as a horizontal red line. Only $k$ values up to $0.1$~Mpc$^{-1}$ are considered, to mitigate the issue of bias in the LR simulations potentially inflating the HR variance estimates. For details, see text (Section~\ref{sec:smooth}).\label{fig:dpc_var}}
\end{figure}

We treat each realization of the spectrum as a draw from a multivariate Gaussian distribution with mean given by an unknown smooth cosmology- and redshift-specific spectrum $\mathcal{P}^{c,z}$, where $c$ denotes the cosmology and $z$ denotes the redshift, modeled by a two-layer process convolution. Thus, for a given series $s \in \{\mbox{PT}, \mbox{LR}, \mbox{HR}\}$, and a given replicate $i = 1,
\cdots, N_s$ (where $N_s$ is the number of simulations for the series, specifically $N_{\text{PT}}=1,$ $N_{\text{HR}}=1,$ and $N_{\text{LR}}=16$), we have a multivariate Gaussian density for the realized
spectrum $P^{c,z}_{s,i}$,
\begin{eqnarray}
\label{eq:smooth_pc_1}
&&f(P^{c,z}_{s,i}) \propto \left| A_s \Omega A_s^{T} \right|^{1/2} \\
&&\times  \exp \left\{ -\frac{1}{2} \left(P^{c,z}_{s,i} - A_s
\mathcal{P}^{c,z} \right)^{T} A_s \Omega A_s^{T}
\left(P^{c,z}_{s,i} - A_s \mathcal{P}^{c,z} \right) \right\}. \nonumber
\end{eqnarray}
Here, $A_s$ is a projection matrix of zeros and ones that is used to remove the low-$k$ values for which only the PT series is used, and the high-$k$ values for which only the HR series is used; and $\Omega_s$ is a diagonal matrix of the precisions (inverse variances), with $\Omega_{PT}$ set to $10^{12}$ to reflect that the PT runs are non-stochastic.

We estimate the variance of the LR and HR simulations (i.e., the inverse of the diagonals of $\Omega_{LR}$ and $\Omega_{HR}$) at each $k$ using information from all cosmologies and redshifts and treat these values as fixed during the smoothing. Specifically, the replicate LR simulations provide cosmology- and redshift-specific estimates of variance for the LR simulations. We model these variances as a function of $k$ via a log-log regression model. The HR simulations should in theory have a variance proportional to that of the LR simulations -- we learn the multiple by which to adjust the LR simulation variance estimates to approximate the HR variance by 1) Using the average of the LR simulations as the mean of the HR simulations; 2) Generating a LR variance estimate from across cosmologies/redshifts for each $k$; 3) Finding the average ratio of HR variance to LR variance across all $k$, and; 4) Letting the HR variance for a given $k$ be equal to the LR variance at that $k$ times this learned multiplier. Figure \ref{fig:dpc_var} shows the results of the log-log regression model for LR variance and the ratio of HR to LR variances along with the ratio average 0.268, i.e. the learned multiplier.

\begin{figure}
\includegraphics[width=\columnwidth]{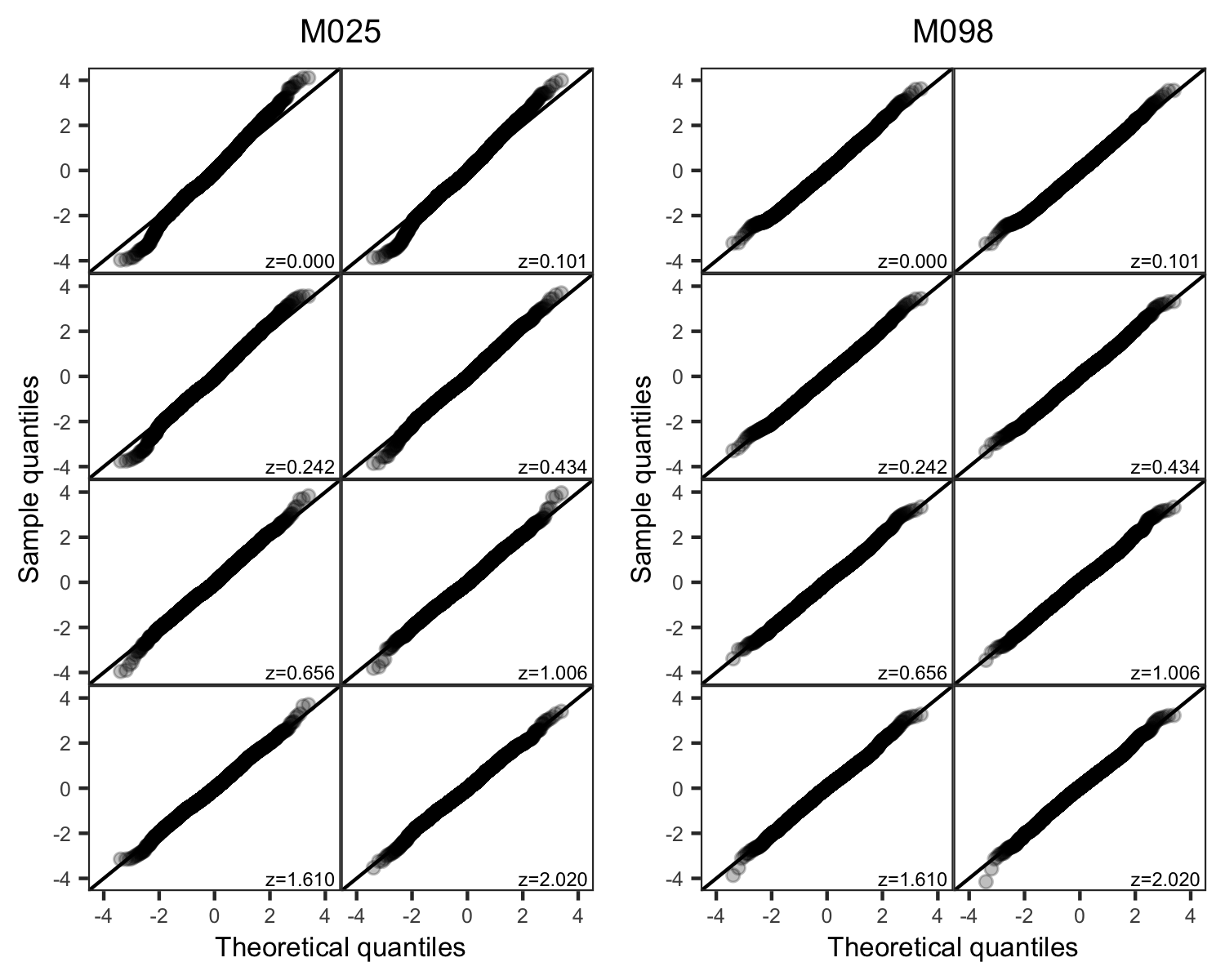}
\caption{Quantile-quantile plots of the standardized residuals for the lower- and higher-resolution simulations at the eight redshifts for two cosmologies. Standardized residuals are computed by subtracting the estimated mean from the simulations and multiplying each value by the square root of the resolution-dependent precision at its $k$ value. We expect the resulting sample to follow a standard normal distribution with no dependence on $k$. The sample quantiles of the standardized residuals are plotted against the theoretical quantiles for a sample of the same size from the standard normal distribution. The tails diverging from the $x=y$ line indicate slightly heavier tails than expected from normality, but overall the model fits relatively well.}
\label{fig:diagnostics_qq_bothres_red}
\end{figure}

The first layer of the process convolution for the smooth power spectrum $\mathcal{P}^{c,z}$ is built via a smoothing matrix $K^{\sigma}$ operating on Brownian motion $u^{c,z}$, i.e., 
\begin{equation} \label{eq:procsmooth}
\mathcal{P}^{c,z} = K^{\sigma} u^{c,z}.
\end{equation}
The Brownian motion vector $u^{c,z}$ has marginal variance $\tau^2_u$ and is realized on a sparse
grid (relative to the power spectrum), $x$, of $M_u=150$ evenly spaced points, i.e., 
\begin{equation} \label{eq:uprior}
f(u^{c,z}) \propto \left| \frac{1}{\tau^2_u} W \right|^{1/2} \exp
\left\{ -\frac{u^{c \prime} W u^{c,z}}{2 \tau^2_u} \right\},
\end{equation}
where $W$ is the $M_u \times M_u$ Brownian precision matrix with diagonal equal to
$[1~2~\cdots~2~1]$ and $-1$ on the first off-diagonals.
The smoothing matrix $K^{\sigma}$ is built using Gaussian kernels
whose width varies smoothly across the domain.  Thus, we have
\begin{equation}
\label{eq:ksigma}
K^{\sigma}_{i,j} = \frac{1}{\sqrt{2 \pi \sigma^{2}_{i}}} \exp \left\{
-\frac{ \left( \log_{10}(k_i) - x_j \right)^2}{2 \sigma_i^2} \right\},
\end{equation}
where $k_i$ is the $i$-th value of $k$ for which the power spectrum is computed and $x_j$ is the $j$-th value of the grid over which the Brownian motion is realized.

In Eqn.~(\ref{eq:ksigma}), $\sigma$ is indexed by $i$ to indicate that it is domain-dependent, i.e. it varies with $k$.  Intuitively, we want to enable $\sigma$ to be small enough in the middle of the domain to capture the oscillations,
but large enough elsewhere to smooth away the noise. We model this varying bandwidth parameter via the second layer of our process
convolution model.
Specifically, the process convolution for the smooth bandwidth parameter, $\sigma$, is built via a smoothing matrix $K^{\delta}$ operating on {\em i.i.d.} Gaussian
variates $v$, i.e., \begin{equation}
\sigma = K^{\delta} v.
\end{equation}
The variates, $v$, have mean zero and variance, $\tau^2_v$, and are observed on an even sparser
grid (relative to $x$) $t$ of $M_v=15$ evenly spaced points,
\begin{equation} \label{eq:vprior}
f(v) \propto \left( \frac{1}{\sqrt{ \tau_{v}^{2} } } \right)^{M_v}
\exp \left\{ -\frac{v^{\prime} v}{2 \tau^2_v} \right\}.
\end{equation}
This smoothing matrix $K^{\delta}$, like $K^{\sigma}$, is built using Gaussian smoothing kernels but with a constant, rather than domain-dependent, bandwidth $\delta$, i.e., 
\begin{equation}
K^{\delta}_{i,j} = \frac{1}{\sqrt{2 \pi \delta^{2}}} \exp \left\{
  -\frac{\left(x_i - t_j  \right)^2}{2 \delta^2}  \right\}.
\end{equation}

Restating the model for spectrum realizations defined in Eqn.~(\ref{eq:smooth_pc_1}) using the process convolution representation of $\mathcal{P}^{c,z}$ from Eqn.~(\ref{eq:procsmooth}), we get a distribution for the simulated power
spectra for a given cosmology and redshift, 
\begin{eqnarray} \label{eq:likelihood}
&&f(P^{c}_{s,i}) \propto  \left| A_s \Omega A_s^{T}\right|^{1/2}\\  
&&\times  \exp \left\{ -\frac{1}{2}
\left(P^{c}_{s,i} - A_s K^{\sigma} u^{c,z} \right)^{T} A_s \Omega
A_s^{T} \left(P^{c}_{s,i} - A_s K^{\sigma} u^{c,z} \right)
\right\}, \nonumber
\end{eqnarray}
where $K^{\sigma}$ depends on the parameters $v$ and
$\delta$.  We choose noninformative inverse gamma, inverse gamma, and uniform priors for $\tau^{2}_{u}$,
$\tau^2_v$, and $\delta$, respectively:
\begin{eqnarray} \label{eq:priors}
\pi(\tau^2_u) & \propto & \left( \frac{1}{\tau^2_u} \right)^2 \exp
\left(-\frac{0.001}{\tau^2_u} \right),  \nonumber \\
\pi(\tau^2_v) & \propto & \left( \frac{1}{\tau^2_v} \right)^2 \exp
\left(-\frac{0.001}{\tau^2_v} \right),  \\
\pi(\delta) & \propto & I\{ 0 \leq \delta \leq R_{\delta} \}, \nonumber
\end{eqnarray}
where $R_{\delta}$ is the maximum of the realized spectra locations.

\begin{figure}
\includegraphics[width=\columnwidth]{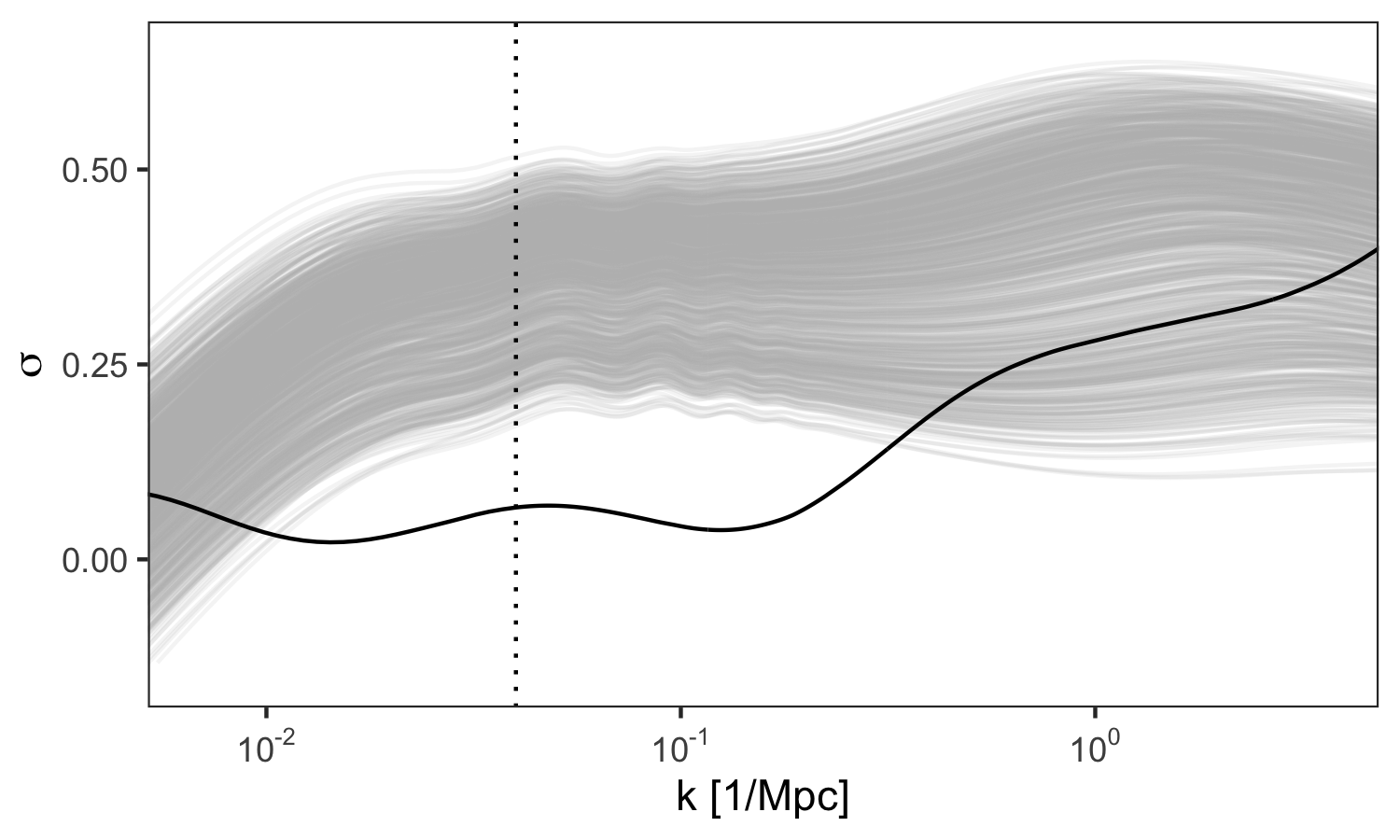}
\caption{Median MCMC draw for the bandwidth function $\sigma$ (black line) in the two-layer process convolution used to estimate the smooth underlying power spectrum (Section~\ref{sec:smooth}). Small values for $\sigma$ correspond to places where the spectra are comparatively less smooth. In addition, the power spectra are shown (111 models, 8 redshifts each) in light gray, scaled to fit the range of $\sigma$. Note that $\sigma$ is smallest on the baryon acoustic oscillation scale, and highest at large $k$, as expected. The vertical line shows the matching point between perturbation theory and the simulation outputs.\label{fig:dpc_sigma}}
\end{figure}

While we are unable to compare our smoothed power spectra to a known truth, we can examine some diagnostics of our modeling assumptions. We assume that observations from the simulated spectra on the emulation scale are independent and normally distributed about a smooth mean with a $k$-dependent variance. Therefore, we compute standardized residuals (i.e., we subtract the smooth mean from the simulations and divide this value by the standard deviation), which we expect to look like independent and identically distributed normal variates with mean 0 and standard deviation 1. We provide quantile-quantile plots of these standardized residuals in Figure \ref{fig:diagnostics_qq_bothres_red}. An approximate match to the $45^{\circ}$ line by the points indicates a good fit. Analogous plots broken down into LR-only and HR-only residuals, as well as plots of these standardized residuals against $k$, are included in the supplementary materials. 

Figure \ref{fig:dpc_sigma} shows the kernel width function $\sigma$ as estimated by the MCMC process via the Metropolis-Hastings algorithm for the two-level process convolution. As expected, $\sigma$ is smallest on the baryon acoustic oscillation scale (i.e., in the vicinity of the baryon wiggles) and biggest at large $k$. In regions where $\sigma$ is smaller, the local values of the latent process receive large weights and the contributions for values further away drop off more quickly. That is, local bumps are not smoothed out. Where $\sigma$ is larger, the contributions of farther away values of the latent process receive more weight -- leading to more smoothing.

\subsection{Emulating the Smooth Power Spectra}

Using the smooth estimates of the matter power spectra, we now build an emulator to predict nonlinear matter power spectra at arbitrary cosmologies within the ranges specified in Eqn.~(\ref{eqn:paramranges}). As in \cite{coyote1}, we build the emulator using an orthogonal basis representation of the spectra. In brief, we first join the smooth spectra across the eight output redshifts so that the length of the observation vector for each of the 111 training cosmologies is $351 \times 8$, i.e. the length of the spectra at a single redshift (size of \{$k$\}) times the number of redshifts. 

We model the 111 joined power spectra
using an $n_\mathcal{P}$-dimensional basis representation: 
\begin{equation}
\mathcal{P}(k,z;\theta)=
\sum_{i=1}^{n_\mathcal{P}}\phi_i(k;z)w_i(\theta), 
\end{equation}
where $\theta$ represents the 8-dimensional vector of cosmological parameters, the $\phi_i(k;z)$ are the basis functions, and the $w_i(\theta)$ are the corresponding weights. The dimensionality $n_\mathcal{P}$ refers to the number of
orthogonal basis vectors
$\{\phi_i(k;z),\dots,\phi_{n_\mathcal{P}}(k;z)\}$. Here $n_\theta=8$; we set $n_\mathcal{P}=45$, which is the number of basis vectors $\phi_i(k;z)$ at which the test and train errors stabilize (i.e., at which further principal components do not add meaningful information to the model). The parameters $\{\theta_1, \ldots, \theta_8 \}$ are directly the cosmological parameters ordered and scaled as specified in Eqn.~(\ref{eqn:paramranges}), with the exception of 
$\theta_7,$ which is defined as $(-w_0-w_a)^{1/4},$ with $0.3\le \theta_7 \le 1.29$.

We use principal components to construct an appropriate set of orthogonal basis vectors $\phi(_i(k;z)$. We model the
weights $w_i(\theta)$ using Gaussian Process (GP) models. GPs are a flexible class of nonparametric models representing
functions that change smoothly with parameter variation, e.g., the
variation of the power spectrum as a function of cosmological
parameters.
We use SEPIA (\citealt{sepia}) to implement both tasks, that is constructing the bases and sampling the GP model parameters and realizations via MCMC. In the MCMC sampler, step sizes are automatically tuned using the YADAS approach (\citealt{sepia}) prior to sampling. We generate 10,000 samples, with the first 2,500 discarded as burn-in. Finally, we predict the spectra for test and training cosmologies using the posterior means of the process convolution parameters from the sampler.

To facilitate its use by the broader research community, we are releasing the fully trained emulator with this paper\footnote{https://github.com/lanl/CosmicEmu}. The emulator outputs nonlinear power spectra at a set of redshifts between $z=0$ and $z=2.02$ out to $k=5~$Mpc$^{-1}$, for any target cosmological model that lies within the parameter bounds specified in Eqn.~(\ref{eqn:paramranges}).

\section{Emulator Performance}
\label{sec:emperf}
In this section we evaluate the emulator performance in several different ways. First, we compare the emulator results to those from a set of additional simulations, each carried out with the same settings as in the Mira-Titan Universe suite (i.e., PT, LR, and HR runs to generate a smooth power spectrum). This is followed by a hold-out (``leave-out-one'') test over five cosmologies that are closest to the center of the sampling design. Next, we show comparisons to alternative fitting/emulation approaches adopted by other groups. Finally, we present comparisons against HACC simulations carried out at significantly higher resolutions than in the Mira-Titan runs. 

\subsection{Comparison to Additional Simulations and Internal Tests}
\label{sec:emu-perform}

\begin{table*}
\begin{center}
\caption{Additional Models for Testing\label{app:tab5}}
\begin{tabular}{lcccccccc}
Model & $\omega_m$  & $\omega_b$ & $\sigma_8$ & $h$  & $n_s$ & $w_0$ & $w_a$ & $\omega_\nu$ \\
\hline\hline
T001 & 0.1433 & 0.02228 & 0.8389 & 0.7822 & 0.9667 & -0.8000 & -0.0111 & 0.008078\\
T002 & 0.1333 & 0.02170 & 0.8233 & 0.7444 & 0.9778 & -1.1560 & -1.1220 & 0.005311\\
T003 & 0.1450 & 0.02184 & 0.8078 & 0.6689 & 0.9000 & -0.9333 & -0.5667 & 0.003467\\
T004 & 0.1417 & 0.02300 & 0.7767 & 0.7256 & 0.9222 & -0.8444 & 0.8222 & 0.004389\\
T005 & 0.1317 & 0.02242 & 0.7611 & 0.6878 & 0.9333 & -1.2000 & -0.2889 & 0.001622\\
M000 & 0.1335 & 0.02258 & 0.8 & 0.71 & 0.963 & -1.0 & 0.0 & 0.0\\
\hline\hline
\end{tabular}
\end{center}
\end{table*}

\begin{figure}
\includegraphics[width=\columnwidth]{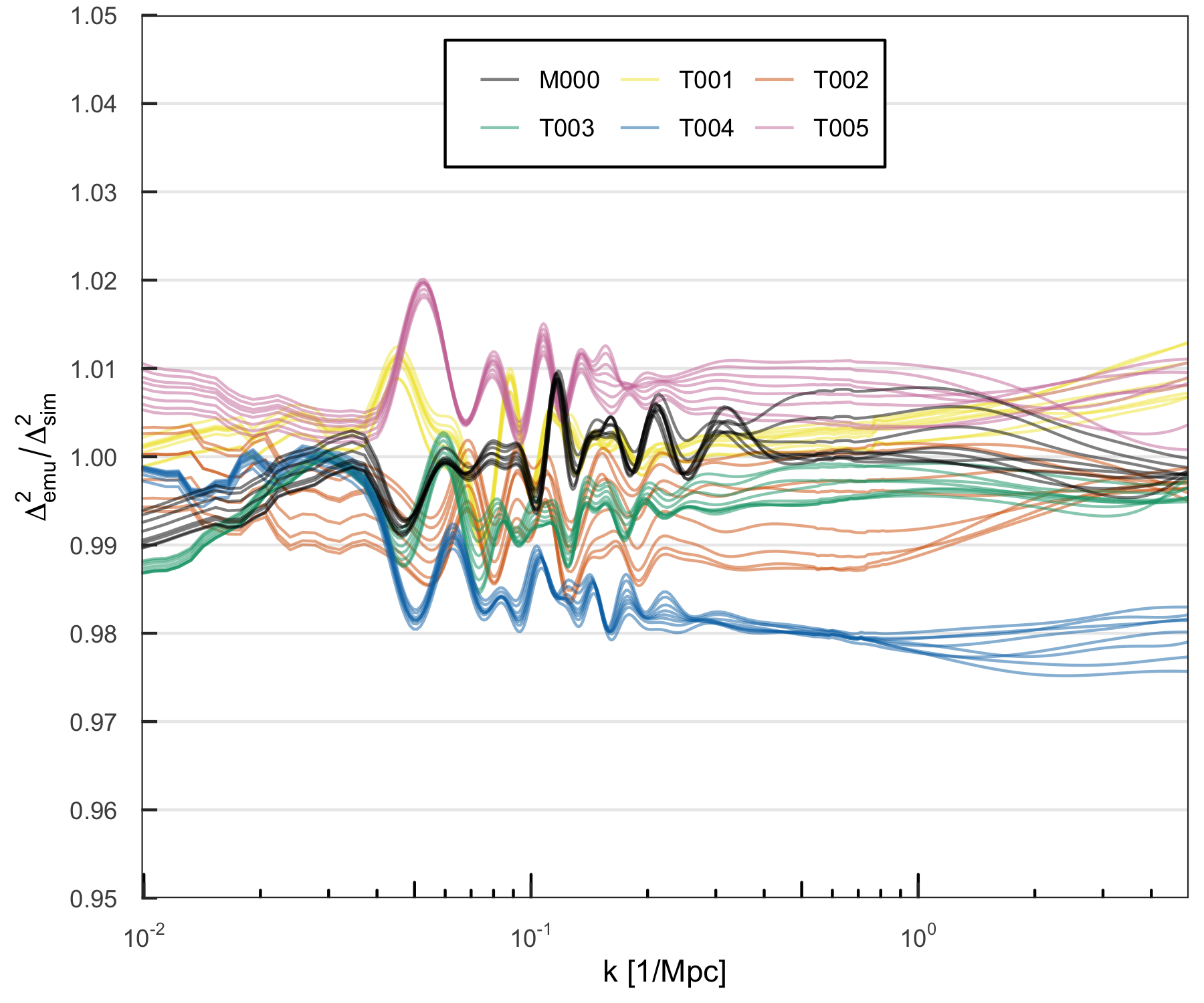}
\caption{Test of the final Mira-Titan emulator based on 111 models for $P_{tot}$, the total matter power spectrum. The comparison is against smooth simulation results from models specified in Table~\ref{app:tab5}, which were constructed in the same way as the model outputs used to build the emulator, but at different values of cosmological parameters (Section~\ref{sec:emu-perform}). Different lines within one color are results for different redshifts. For all but T004 (a model very close to the edge of the parameter hypercube), which has an accuracy of 2.48\%, the overall accuracy is better than 2\%.\label{fig:ptot}}
\end{figure}

\begin{figure}
\includegraphics[width=\columnwidth]{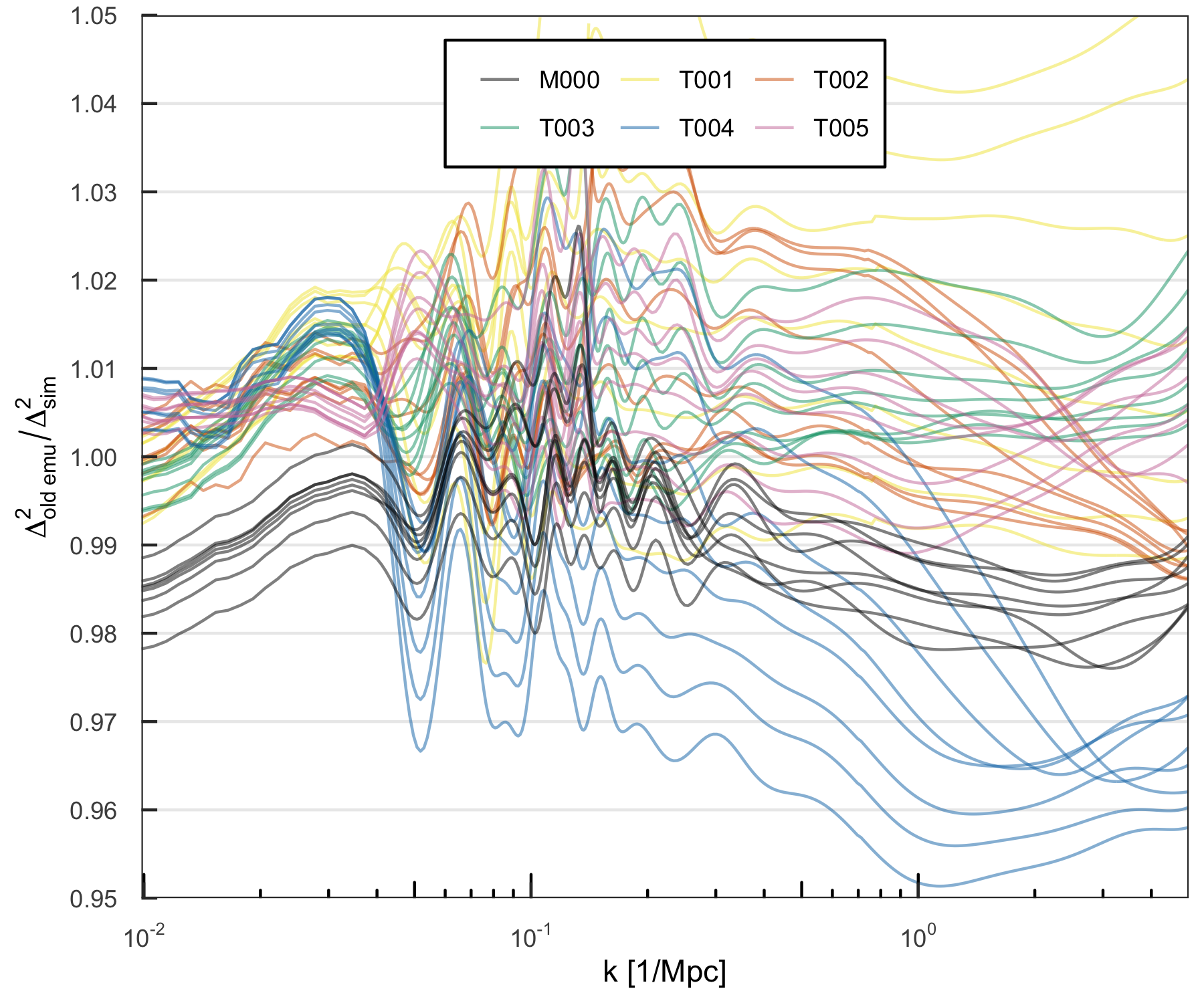}
\caption{Relative comparison against the current emulator results for $P_{tot}$ (Figure~\ref{fig:ptot}) versus the emulator built on the first set of Mira-Titan simulations based on 36 cosmological models \citep{MT-pow1}. The older emulator is accurate at the 3-6\% level; specifically, the test runs shown have ratio $\Delta^2_{\text{emu}} / \Delta^2_{\text{sim}}$ ranging from 0.951 to 1.061.\label{fig:ptot_old}}
\end{figure}

In order to carefully test the emulator performance, we carried out five additional simulations (`T models') that span the full cosmological parameter space of the Mira-Titan Universe and one $\Lambda$CDM model, representing the best fit WMAP-7 cosmology~\citep{wmap7} (we refer to this model in the following as WMAP7 M000). These six simulations are not part of the original design. The simulation specifications with regard to volume and resolution and the number of realizations for the LR simulations are the same as for the 111 simulations as well as the smoothing strategy to obtain the final power spectra. This allows for a direct comparison with the emulator without having to take into account issues such as realization dependence of individual simulations. The cosmologies for the simulations are listed in Table~\ref{app:tab5}. Figure \ref{fig:ptot} shows tests of the emulator accuracy for $P_{tot}$, the total matter power spectrum between $z=0$ and $z=2.02$. 
The final Mira-Titan emulator based on 111 models has an accuracy better than 2\% for all but one of the test models. The best out-of-sample performer is M000, the $\Lambda$CDM model, with accuracy at or below 1\% for all $k$. The worst out-of-sample performer is T004 which has an accuracy at or below 2\% at low to medium $k$ and 2.48\% at higher $k$. For this cosmology, $w_0 + w_a = -0.0222$ which is both extremely close to the edge of the parameter space and nearly unphysical. As such, this represents an extreme test of the emulator and the performance is still very acceptable. By comparison, the emulator from \citep{MT-pow1} that is based on the first set of Mira-Titan simulations for 36 cosmological models is accurate at the 3-6\% level, shown in Figure \ref{fig:ptot_old}.

\begin{figure}
\includegraphics[width=\columnwidth]{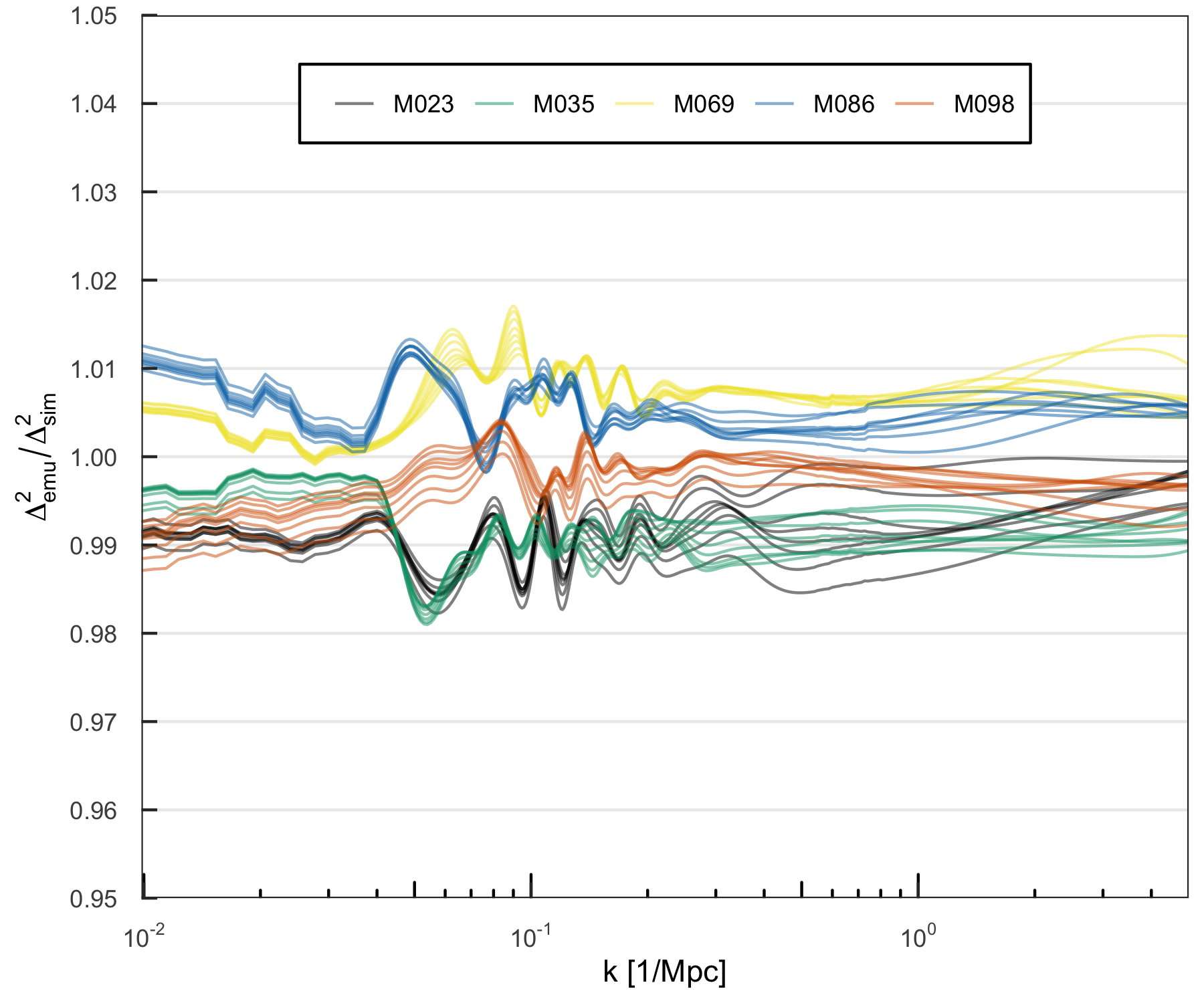}
\caption{Hold-out test (leave-out-one) cross-validation results for the $P_{tot}$ emulator on the five cosmologies closest to the center of the design. Each of the test cosmologies has been left out of the emulator construction for the relevant individual comparison; different lines within one color are results for different redshifts.\label{fig:perf_crossval}}
\end{figure}

Figure \ref{fig:perf_crossval} shows hold-out (leave-out-one) cross-validation results for the $P_{tot}$ emulator on the five cosmologies closest to the center of the design. Accuracy is around or below 2\% for all cosmologies. In this comparison, the emulator is constructed by leaving out one of the models, results from which are then used to test the quality of the emulator. Because information is lost during the construction of the ``leave-out-one'' emulator, this error is in principle larger than for the emulator constructed using all the models in the sampling design.

In-sample predictions for $P_{tot}$ -- predictions for the training set from the emulator -- with the new emulator exhibit similar ``worst case'' errors to the out-of-sample observations. That is, the ratio $\Delta^2_{\text{emu}} / \Delta^2_{\text{sim}}$ is at worst 0.977 across $k$ for the fringe in-sample cosmologies (compare to 0.975 for the test cosmology T004). A visual for all cosmologies is shown in the supplementary materials. A comparison with Figure 6 of the Mira-Titan II paper (\citealt{MT-pow1}) shows substantial improvement on in-sample prediction. (Note that an emulator is typically expected to interpolate on the training set exactly, but our methodology behaves somewhat more like regression, where one does not interpolate, but aims to minimize the error.)

Note that in all performance plots we use ``emu'' or ``CosmicEmu'' to refer to the emulator built using the complete set of Mira-Titan IV simulations. Analogous figures for $P_{cb}$, the CDM and baryon component of the matter power spectrum, are shown in the supplementary materials. The results are very similar to those for $P_{tot}$.

\subsection{Comparison with Alternative Approaches}


As discussed in Section~\ref{sec:intro}, the importance of fast prediction schemes for the power spectrum has been long recognized and different approaches have been developed to address this challenge. These approaches employ different strategies, from fitting functions to emulation approaches broadly similar in spirit to those implemented here. In this section, we compare the accuracy of our new emulator to four  prediction schemes (three of which have been recently released) that are commonly used for analysis and forecasting tasks. The four prediction tools we chose for comparison are all based on somewhat different construction strategies than those used here and therefore provide the opportunity to compare and contrast different methods, as well as test for the accuracies achieved. 

\begin{figure}
\includegraphics[width=\columnwidth]{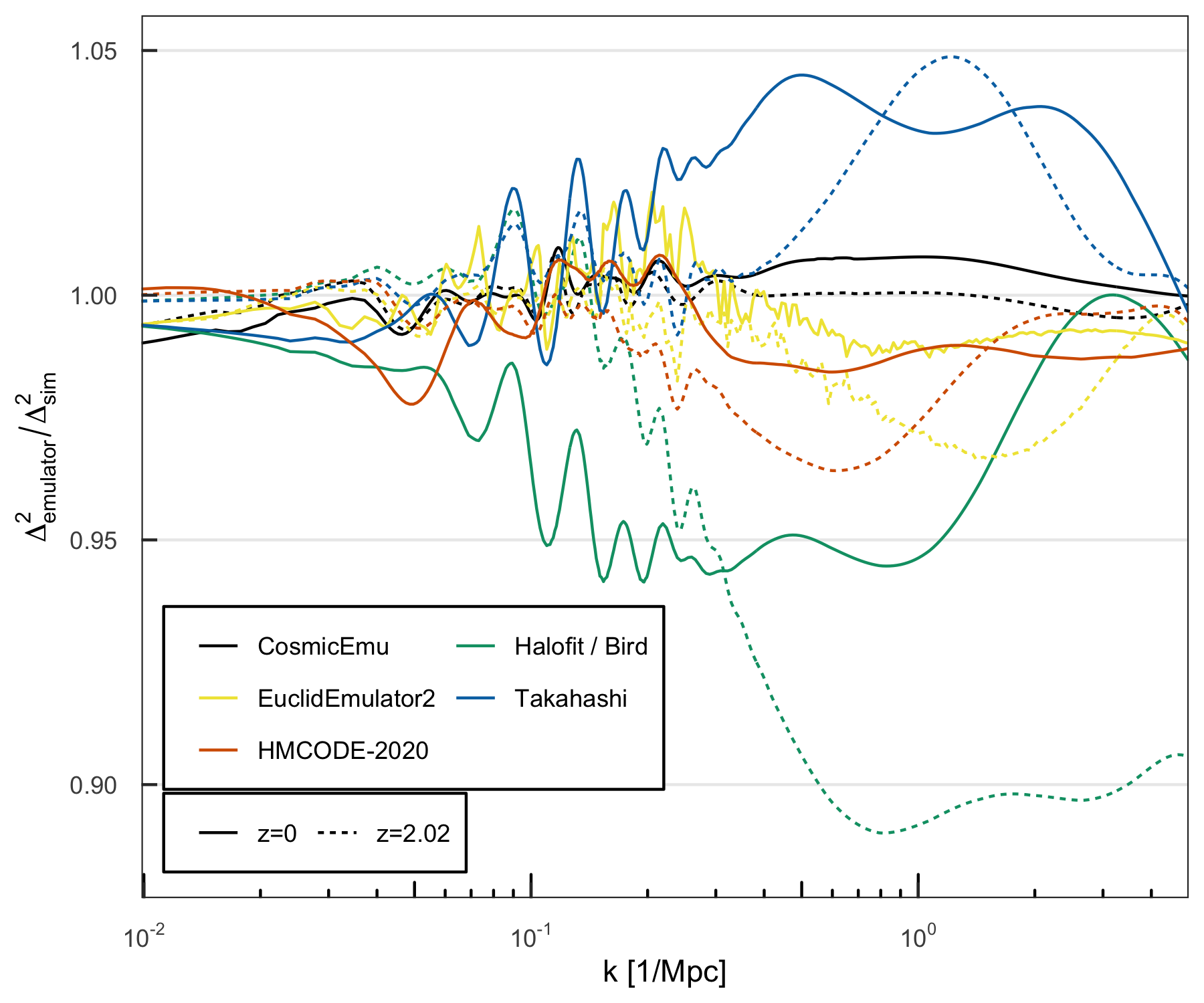}
\caption{Comparison of different prediction methods relative to the $P_{tot}$ simulation result for $z=0$ and $z=2.02$ on the WMAP7 M000 cosmology. Results from our emulator, the CosmicEmu, are shown in black. Exact agreement with the simulation would result in a ratio of unity.\label{fig:perf_comp_m000}}
\end{figure}

Our choice was also guided by the desire to compare the predictions covering the full range of cosmological parameters introduced in this paper, including massive neutrinos and a varying dark energy equation of state. In our previous work on the Mira-Titan power spectrum emulator \citep{MT-pow1}, fewer alternatives were available and our comparison was necessarily rather restricted. To some extent this is still true, as the parameter ranges we consider are significantly broader than those allowed by alternate methods and the simulations cover cosmologies having nonzero neutrino mass (most alternative methods, including one we include here, do not allow for variations in $w_a$). We show results for the WMAP7 M000 cosmology, which has zero neutrino mass and parameters within the ranges covered by all alternative methods, in Figure~\ref{fig:perf_comp_m000} using all of the different methods. Other prediction schemes exist, e.g., from the BACCO project~\cite{2020MNRAS.499.4905C}. However, their predictions do not cover the full $k$ or redshift range shown in this paper.  
HMCODE-2020 provides predictions over the full parameter range considered in this paper. We are therefore able to show results for all test cosmologies using HMCODE-2020 in Figure~\ref{fig:perf_comp}. In the following we provide a brief summary of the implementations of the different methods and the considered comparisons.

Finally, we also include comparisons of the emulator-predicted spectra against $\Lambda$CDM simulation results from the Last Journey and Farpoint runs~\citep{LJ,Farpoint} undertaken with HACC. Even though the emulator was trained using lower resolution simulations (and lower volume relative to Last Journey), it exhibits very good performance when compared to these significantly higher resolution simulations.

\subsubsection{Halofit}

The Halofit approach has a relatively long history. We use two versions in this paper for comparison, first, the approach as originally implemented by~\cite{smith03} and extended by \cite{bird} to include neutrinos and to provide a better fit on small scales.  The second version we compare to was developed by~\cite{takahashi12} who refined the fit further. 
The work by~\cite{smith03} is based on a set of scale-free and CDM N-body simulations that they use to derive a fitting formula for the power spectrum inspired by the halo model; the reported accuracy is at the 7\% level. \cite{takahashi12} carried out a new set of sixteen simulations, some based on the results from the WMAP survey, some following the simulation design strategy used in~\cite{coyote3}, and extended the number of fitting parameters to 35 from the original 30 parameters in~\cite{smith03}.  They cover a wide $k$ and redshift range and report an accuracy at the 5-10\% level, depending on the $k$-range covered. \cite{bird} carried out a set of simulations including massive neutrinos. They then extended the~\cite{smith03} Halofit to allow for predictions of the effect of massive neutrinos on the nonlinear matter power spectrum. 

We provide a comparison for two Halofit implementations, \cite{takahashi12} and \cite{bird}, for the $\Lambda$CDM simulation WMAP-7 M000 at two redshifts. Results are shown in Figure~\ref{fig:perf_comp_m000}. The Takahashi implementation has a worst-case error of close to 5\%, while the Halofit/Bird implementation has a worst-case error of $\sim$11\%. We also show in the same figure the emulator-predicted WMAP-7 M000 spectra at the same two redshifts.

Our comparison is based on the publicly available version implemented in CAMB~\citep{camb}, version~1.3.2. The Halofit implementations were not originally designed to allow for the variation of $w_a$, therefore we do not carry out a direct comparison with our additional simulation results from T001-T005.




\subsubsection{EuclidEmulator2}

Next, we compare our approach to the recently released EuclidEmulator2, described in~\cite{2021MNRAS.505.2840E}. As for Halofit, the EuclidEmulator2 has been improved over time with the first implementation described in~\cite{2019MNRAS.484.5509E}. The approach taken in this work uses a supervised regression technique, the polynomial chaos expansion (PCE), in combination with a PCA used in a way similar to that reported here. The emulator is based on 250 N-body simulations that each evolve 3000$^3$ particles in volumes of 1($h^{-1}$Gpc)$^3$. Instead of emulating the power spectrum directly, \cite{2021MNRAS.505.2840E} predict the nonlinear correction to the power spectrum. They report that the inclusion of $w_a$ appears to pose a considerable challenge in achieving high accuracy predictions. 

While variations in $w_a$ \textit{are} allowed by EuclidEmulator2, the cosmological parameter bounds they support are narrower than ours and M000 is the only test cosmology listed in Table~\ref{app:tab5} to fall within the parameter ranges specified by the EuclidEmulator2 (note that we are treating dark energy purely as a background contribution, whereas~\citealt{2021MNRAS.505.2840E} also allow for dark energy perturbations). We show the comparison of the EuclidEmulator2 predictions to our simulation results for WMAP-7~M000 in Figure~\ref{fig:perf_comp_m000}. The EuclidEmulator2 has a worst-case error of $\sim$3\%.

\subsubsection{HMCODE-2020}

The HMCODE-2020, described in~\cite{2021MNRAS.502.1401M}, has also already undergone several updates. First introduced in~\cite{mead16}, the most recent version allows for the inclusion of baryonic effects and improved predictions for neutrinos. HMCODE-2020 uses the halo model idea as a starting point and then adds parameters to provide flexibility to match nonlinear power spectrum results from simulations. 

\begin{figure}
\includegraphics[width=\columnwidth]{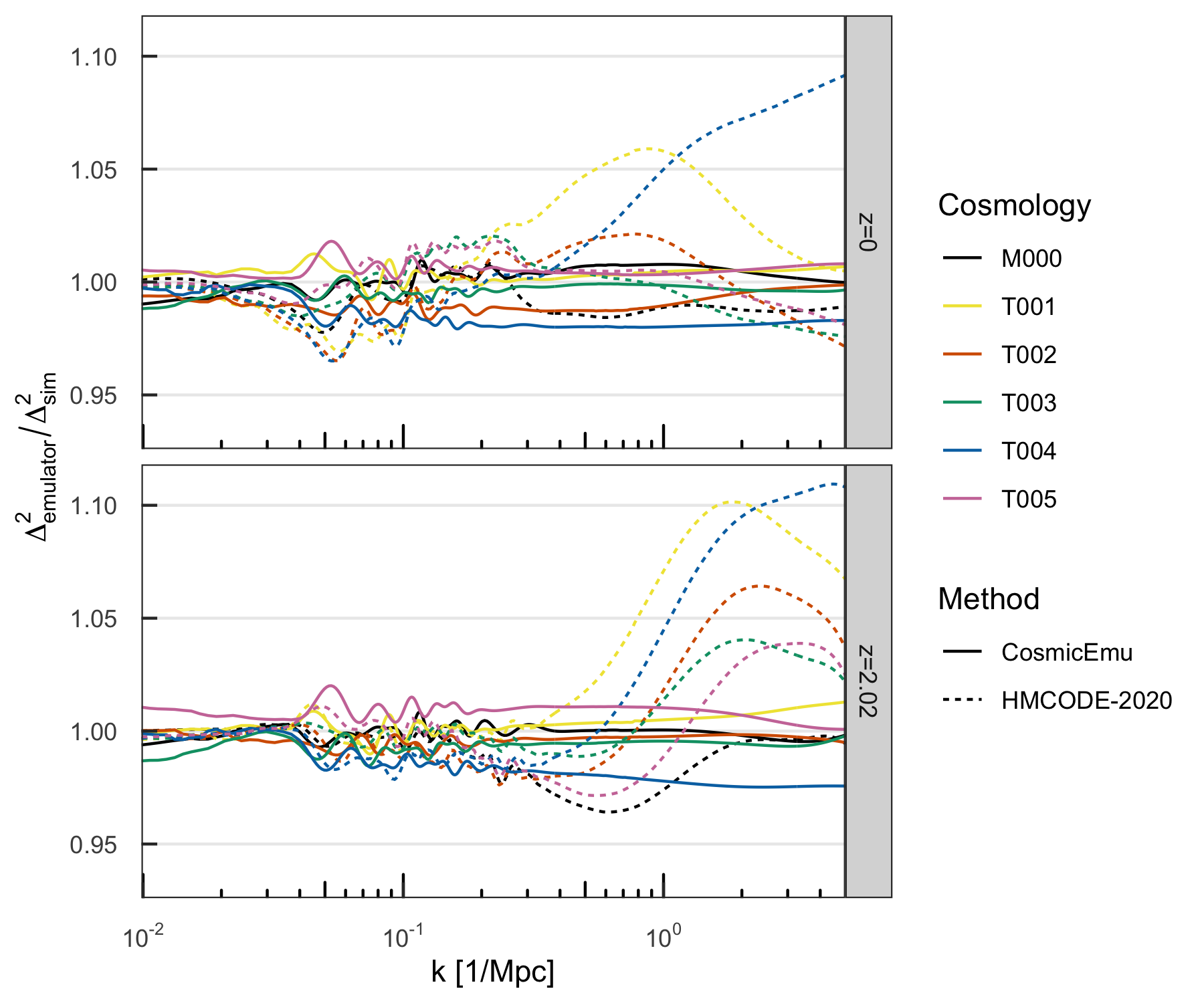}
\caption{Relative error of the HMCODE-2020 predictions in predicting the $P_{tot}$ simulation results for $z=0$ and $z=2.02$ on the test cosmologies (dashed lines), with the analogous results for the CosmicEmu (solid lines) shown for comparison.\label{fig:perf_comp}}
\end{figure}

As with the Halofit comparison, we generate HMCODE-2020 predictions with CAMB, version~1.3.2. Since HMCODE-2020 spans a broad range of cosmological parameters and allows for variations in $w_a$, we present a comparison to our simulation results for M000 and T001-T005 in Figure \ref{fig:perf_comp}. Our emulator outperforms HMCODE-2020 in terms of absolute error integrated over logged $k$ for all test cosmologies and redshifts; the supplemental materials include additional details on this comparison. Note the HMCODE-2020 results for M000 are also included in Figure \ref{fig:perf_comp_m000}. The HMCODE-2020 predictions have a worst-case error of 11\% (occurring for T004 -- as also the case with our emulator).

\subsection{Comparison to Other Simulations}

\begin{figure}
\centering
\includegraphics[width=\columnwidth]{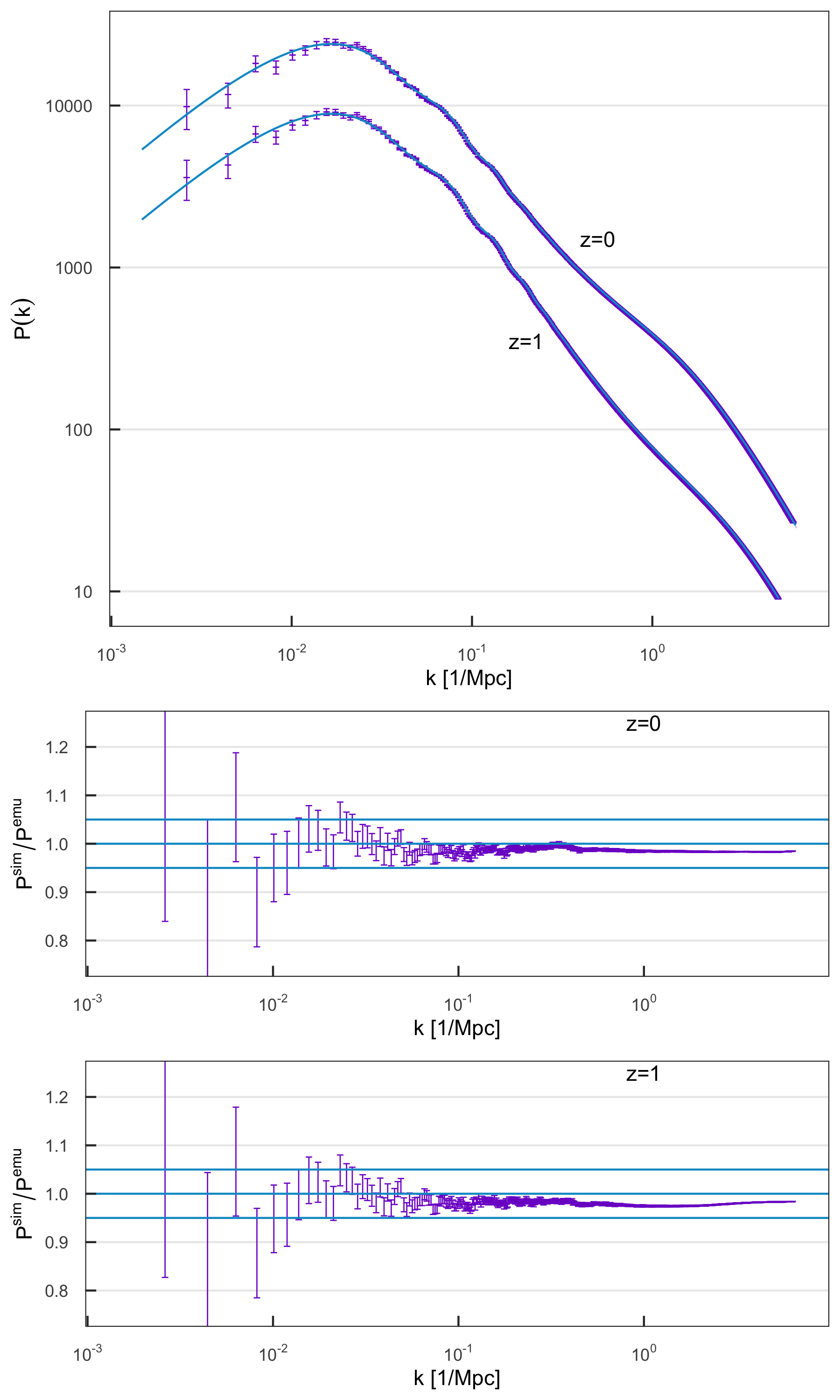}
\caption{Comparison of the Last Journey simulation to the emulator output for the Planck M000 cosmology at $z=0$ and $z=1$. The horizontal lines represent $\pm 5$\% error limits. \label{fig:supplement_sim_lj}}
\end{figure}

\begin{figure}
\centering
\includegraphics[width=\columnwidth]{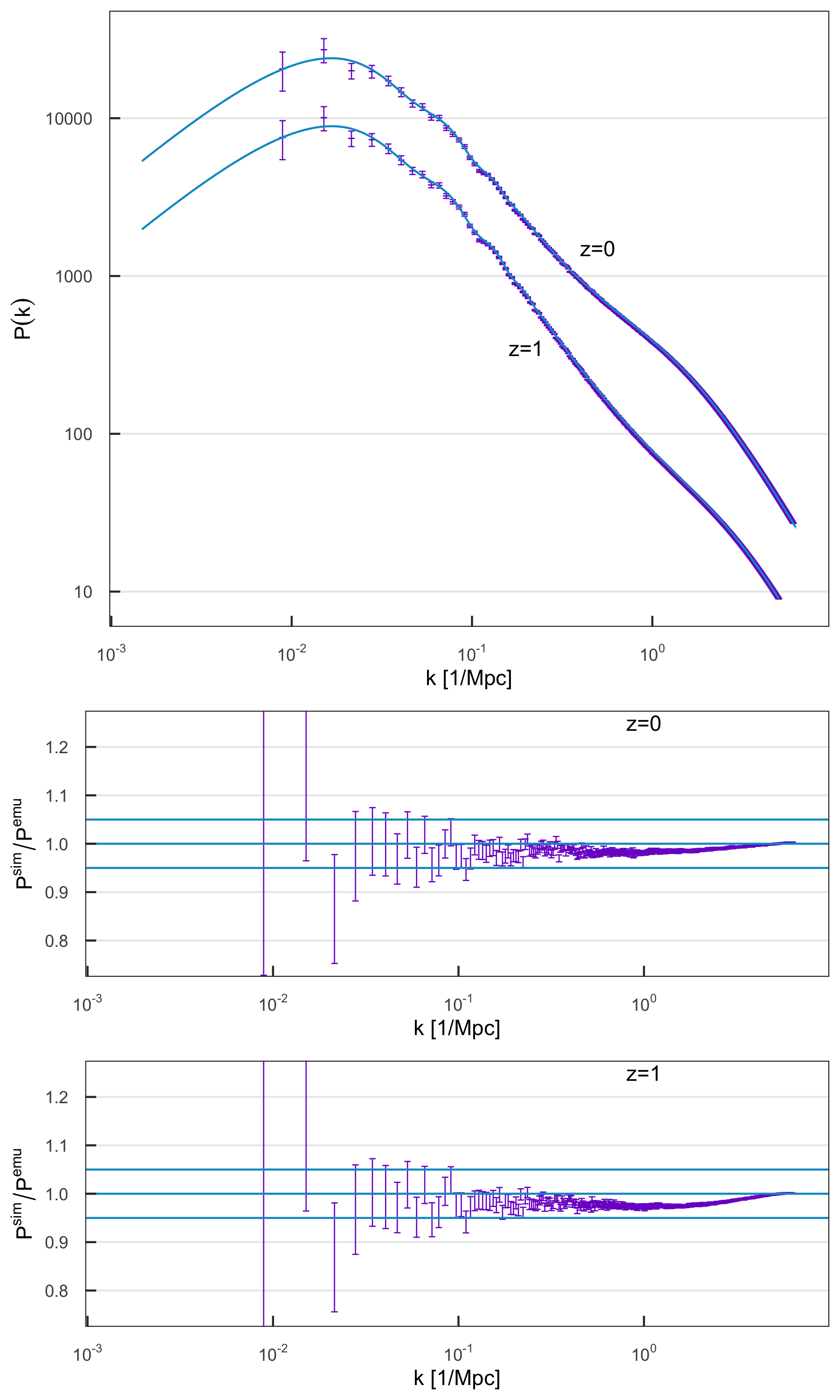}
\caption{Comparison of the Farpoint simulation to the emulator output for the Planck M000 cosmology at $z=0$ and $z=1$.\label{fig:supplement_sim_fp}}
\end{figure}

In Section~\ref{sec:emu-perform} we compared the emulator to a set of additional simulations that were set up in the same way as the Mira-Titan suite with regard to volume, resolution, and data preparation. However, the natural question is whether the emulator will also perform well when tested against higher-resolution simulations. In \cite{coyote1} we presented detailed tests to establish simulation requirements. In particular, the simulation volume and/or mass resolution are of major concern. To address these potential concerns, we include comparisons of the emulator output for two additional simulations, the Last Journey and Farpoint runs. The Last Journey simulation covers a volume of (5.25Gpc)$^3$ with a mass resolution of $m_p\sim 4\cdot 10^9$M$_\odot$. The Farpoint simulation covers a volume of (1.48Gpc)$^3$ with a mass resolution of $m_p\sim 7\cdot 10^7$M$_\odot$. Both simulations follow a best-fit Planck cosmology~\cite{planck18}; we call this model Planck M000, indicating that it is a different $\Lambda$CDM model compared to WMAP-7 M000. Figure~\ref{fig:supplement_sim_lj} shows the results for the Last Journey simulation for redshifts $z=0$ and $z=1$ and  
Figure~\ref{fig:supplement_sim_fp} shows the results for the Farpoint simulation for the same two redshifts. The lower panels in each figure show the ratios of the simulations to the emulators at the two redshifts. Overall, the agreement is at the same level as the emulator accuracy reported, indicating that the simulation parameters for the Mira-Titan Universe suite are sufficient for achieving the overall targeted accuracy. Note the appearance of a small bias at higher $k$ relative to the Last Journey simulation, where our emulator slightly over-predicts the simulation result.



\section{Summary and Outlook}
\label{sec:sum}

In this paper we have presented the final Mira-Titan Universe power spectrum emulator, built on the complete set of 111 models from the full simulation suite. This work directly follows from~\cite{MT-pow1} and is the end implementation of the strategy outlined in~\cite{heitmann15}, where we showed that a carefully constructed simulation design can allow for systematic and uniform emulator improvement by adding more simulations to an initially chosen set. In~\cite{MT-pow1} we achieved an accuracy at the six percent level using only 36 cosmological models. Our new emulator, based on 111 cosmological models (including the original 36), provides predictions at the two percent level of accuracy over most of the cosmologies and length scales covered.

The emulator construction is based on the following inputs. For each cosmological model, we generate a smooth prediction (via a two-layer process convolution) by combining results from perturbation theory (PT), a set of lower resolution simulations (LR) and one high resolution simulation (HR). These smooth power spectrum predictions form the basis for constructing the emulator using an empirical principal components decomposition (45 orthogonal basis vectors) with the weights determined by Gaussian process modeling -- the GP model parameters and realizations are sampled using MCMC. 

We presented a range of tests to evaluate the overall accuracy of the emulator, including a set of six additional simulations for cosmologies that were not used in its construction. The emulator provides predictions spanning an eight-dimensional cosmological parameter space and redshifts between $z=0$ and $z=2.02$. The parameter space covered by our emulator is larger than the prediction space provided by most other approaches. For this reason, we were only able to carry out a comparison for the additional simulations beyond $\Lambda$CDM cosmologies with the work of~\cite{2021MNRAS.502.1401M}. Other approaches either did not allow for the variation of all eight parameters or had much narrower coverage of the parameter space. A comprehensive comparison with other methods for a $\Lambda$CDM cosmology showcased the performance of our approach. 

While the accuracy of the current emulator is already very good, it can be improved further by adding additional information on large scales (small $k$ values) as we have shown previously -- the key point is that emulators can be constructed using models that provide information on multiple scales, where each model does not have to be complete in its coverage of the global scales of interest. In \cite{emu_ext} and \cite{MT-pow1}, we augmented the set of full power spectra (i.e., matched PT and simulations) with partial power spectra using perturbation theory alone for select additional cosmologies. Due to increased model discrimination, this augmentation improved the emulator performance not only at low $k$ but across the entire $k$ range, i.e. both above and below the $k$ value at which the PT runs are cut off. Given that the generation of PT predictions is inexpensive, we can follow this approach by adding a large set of PT predictions at new parameter settings specified by our sequential sampling scheme. Addition of this information can tighten the error boundary further to the 1\% level.

A number of precision emulators can be built directly on the baseline of the effort described in this paper, including emulators for correlation functions (density field and halos) and for halo and group/cluster bias at mass scales above $\sim 10^{13}$M$_{\sun}/h$. Several other emulators can be constructed using modeling approaches based on the galaxy-halo connection.

Another of our long-term goals is the robust modeling and inclusion/marginalization of baryonic effects, a key issue in being able to extract cosmological information from observational results at small scales. \cite{2021MNRAS.502.1401M} developed an approach to tackle this challenge based on a halo-model like approach. The method relies on sufficiently accurate hydrodynamic simulation input to tune the modeling parameters. Another approach is to reduce baryon-induced systematics by using a PCA-based mitigation technique~\citep{eifler15} relying on incorporating results from a limited number of hydrodynamical simulations.

A natural extension of the N-body simulation ensembles used for emulation would be to run a comprehensive suite of large volume, sufficiently resolved, hydrodynamical simulations covering both cosmological and baryonic modeling parameters, constructing emulators based on the results. While such a suite will be expensive with regard to computational resources, both due to the extension of the number of parameters and the increase of computational costs compared to gravity-only simulations, it would provide the opportunity to carry out comprehensive studies of baryonic effects across multiple cosmological probes. Such a campaign has to be carefully designed with regard to (subgrid) parameter choices and simulation specifications to guarantee both prediction coverage and a wide range of applicability across the targeted probes. At the same time, careful consideration has to be given to the feasibility of such a campaign in terms of minimizing computational costs. Such an ambitious project would be a natural next step following the Mira-Titan Universe and is currently in planning stages.



\section*{Acknowledgements}
We gratefully recognize the efforts of HACC team members Hal Finkel, Nicholas Frontiere, Vitali Morozov, and Adrian Pope in developing the Mira-Titan simulation suite. KH and SH record their thanks to the Aspen Center for Physics where part of this work was performed, with support provided to the Center by the National Science Foundation grant PHY-1607611. This research used resources of the Argonne Leadership Computing Facility at the Argonne National Laboratory, which is supported by the Office
of Science of the U.S. Department of Energy under Contract
No. DE-AC02-06CH11357. This work also used resources of the Oak Ridge Leadership Computing Facility, which is a DOE Office of Science User Facility supported under Contract DE-AC05-00OR22725. This work was in part supported by the U.S. Department of Energy, Office of Science, Office of Advanced Scientific Computing Research, Scientific Discovery through Advanced Computing (SciDAC) program. Approved for public release: LA-UR-22-27416.

\section*{Data availability}

The code to run the emulator is available at \url{github.com/lanl/CosmicEmu}; feedback from users is welcome. The full cosmological parameter space is included in the Appendix. Simulations underlying this work are from the Mira-Titan campaign, first described in~\cite{heitmann15}; the power spectra from the simulations used to build the emulator will be made available on request.

\begin{appendix}
\label{sec:appendix}
\section{Mira-Titan Universe Model Space}
In this Appendix we provide several tables with the model parameters used in the Mira-Titan Universe simulation suite. The first, Table~\ref{app:tab1}, lists the massless neutrino models. The ten models are based on an independent, complete symmetric Latin hypercube design. The second, Table~\ref{app:tab2}, covers the first segment of our nested design, described in \cite{heitmann15}. Tables~\ref{app:tab1} and \ref{app:tab2} combined were used to generate the first matter power spectrum emulator described in \cite{MT-pow1}. Table~\ref{app:tab3} contains the second segment of the nested design and Table~\ref{app:tab4} completes the full 111 model design, also used in~\cite{MT-massf}. For testing purposes we carried out simulations for five additional cosmologies, following smoothing and joining procedures identical to those for the 111 models used as the basis for the emulator; the parameters for these cosmologies are summarized in Table~\ref{app:tab5}.

\section{Data Alignment Details}

The first alignment step primarily adjusts for the finite size of the LR simulation boxes, which while large (1.3~Gpc to a side), still have an associated $P(k)$ finite-size suppression of about $\sim 0.1$\% at the smallest $k$ values. For each unique cosmology and redshift combination, we
\begin{enumerate}
        \item Calculate $C^{\ell}_{\text{LR}}$, the average of the LR spectra for outputs in the window $k \in [0.022, 0.04]$~Mpc$^{-1}$.
        \item Calculate $C_{\text{PT}}$, the average of the PT for outputs in the window $k \in [0.022, 0.04]$~Mpc$^{-1}$.
        \item Adjust the LR spectra by a constant shift to best agree with the PT results.
    \end{enumerate}
In the second step we align the HR spectra to the LR spectra in the same way as in \cite{emu_ext}. This time, for each unique cosmology and redshift combination, we:
    \begin{enumerate}
        \item Calculate $C_{\text{HR}}$, the average of the HR spectra for outputs in the window $k \in [0.1345, 0.1646]$~Mpc$^{-1}$.     
        \item Calculate $C{\text{LR}}$, the average of the LR spectra for outputs in the window $k \in [0.1345, 0.1646]$~Mpc$^{-1}$.
        \item Adjust the HR spectra by a constant shift to best agree with the LR results.
\end{enumerate}

\section{Corrections to the previous emulator release}
In the process of building the current version, we discovered errors in the previous emulator release~\citep{MT-pow1}. These have been corrected in the new emulator, which fully supersedes the previous version. One was due to an incorrect match of the output time step in the N-body code to the corresponding redshift. This introduced a redshift error of $\Delta z = 0.047$ when matching the power spectra between the N-body simulations and perturbation theory at $z=0.656$. In addition, the growth function with massive neutrinos was normalized at the low-$k$ limit instead of the (preferred) high-$k$ limit where the function approaches a constant. This would have introduced an error of order a percent in the regime where perturbation theory transitions to the lower-resolution N-body power spectra. The results shown in Figure~\ref{fig:ptot_old} are based on the original publicly released emulator described in~\cite{MT-pow1}; no corrections were made to the previous emulator release itself.

\begin{table*}
\begin{center}
\caption{Models with Massless Neutrinos\label{app:tab1}}
\begin{tabular}{lcccccccc}
Model & $\omega_m$  & $\omega_b$ & $\sigma_8$ & $h$  & $n_s$ & $w_0$ & $w_a$ & $\omega_\nu$ \\
\hline\hline
M001 & 0.1472 & 0.02261 & 0.8778 & 0.6167 & 0.9611 & -0.7000 & 0.67220 & 0.0\\
M002 & 0.1356 & 0.02328 & 0.8556 & 0.7500 & 1.0500 & -1.0330 & 0.91110 & 0.0\\
M003 & 0.1550 & 0.02194 & 0.9000 & 0.7167 & 0.8944 & -1.1000 & -0.28330 & 0.0\\
M004 & 0.1239 & 0.02283 & 0.7889 & 0.5833 & 0.8722 & -1.1670 & 1.15000 & 0.0\\
M005 & 0.1433 & 0.02350 & 0.7667 & 0.8500 & 0.9833 & -1.2330 & -0.04445 & 0.0\\
M006 & 0.1317 & 0.02150 & 0.8333 & 0.5500 & 0.9167 & -0.7667 & 0.19440 & 0.0\\
M007 & 0.1511 & 0.02217 & 0.8111 & 0.8167 & 1.0280 & -0.8333 & -1.00000 & 0.0\\
M008 & 0.1200 & 0.02306 & 0.7000 & 0.6833 & 1.0060 & -0.9000 & 0.43330 & 0.0\\
M009 & 0.1394 & 0.02172 & 0.7444 & 0.6500 & 0.8500 & -0.9667 & -0.76110 & 0.0\\
M010 & 0.1278 & 0.02239 & 0.7222 & 0.7833 & 0.9389 & -1.3000 & -0.52220 & 0.0\\
\hline\hline
\end{tabular}
\end{center}
\end{table*}

\begin{table*}
\begin{center}
\caption{First Set of Models from the Nested Design\label{app:tab2}}
\begin{tabular}{lcccccccc}
Model & $\omega_m$  & $\omega_b$ & $\sigma_8$ & $h$  & $n_s$ & $w_0$ & $w_a$ & $\omega_\nu$ \\
\hline\hline
M011 & 0.1227 & 0.0220 & 0.7151 & 0.5827 & 0.9357 & -1.0821 & 1.0646 & 0.000345\\
M012 & 0.1241 & 0.0224 & 0.7472 & 0.8315 & 0.8865 & -1.2325 & -0.7646 & 0.001204\\
M013 & 0.1534 & 0.0232 & 0.8098 & 0.7398 & 0.8706 & -1.2993 & 1.2236 & 0.003770\\
M014 & 0.1215 & 0.0215 & 0.8742 & 0.5894 & 1.0151 & -0.7281 & -0.2088 & 0.001752\\
M015 & 0.1250 & 0.0224 & 0.8881 & 0.6840 & 0.8638 & -1.0134 & 0.0415 & 0.002789\\
M016 & 0.1499 & 0.0223 & 0.7959 & 0.6452 & 1.0219 & -1.0139 & 0.9434 & 0.002734\\
M017 & 0.1206 & 0.0215 & 0.7332 & 0.7370 & 1.0377 & -0.9472 & -0.9897 & 0.000168\\
M018 & 0.1544 & 0.0217 & 0.7982 & 0.6489 & 0.9026 & -0.7091 & 0.6409 & 0.006419\\
M019 & 0.1256 & 0.0222 & 0.8547 & 0.8251 & 1.0265 & -0.9813 & -0.3393 & 0.004673\\
M020 & 0.1514 & 0.0225 & 0.7561 & 0.6827 & 0.9913 & -1.0101 & -0.7778 & 0.009777\\
M021 & 0.1472 & 0.0221 & 0.8475 & 0.6583 & 0.9613 & -0.9111 & -1.5470 & 0.000672\\
M022 & 0.1384 & 0.0231 & 0.8328 & 0.8234 & 0.9739 & -0.9312 & 0.5939 & 0.008239\\
M023 & 0.1334 & 0.0225 & 0.7113 & 0.7352 & 0.9851 & -0.8971 & 0.3247 & 0.003733\\
M024 & 0.1508 & 0.0229 & 0.7002 & 0.7935 & 0.8685 & -1.0322 & 1.0220 & 0.003063\\
M025 & 0.1203 & 0.0230 & 0.8773 & 0.6240 & 0.9279 & -0.8282 & -1.5005 & 0.007024\\
M026 & 0.1224 & 0.0222 & 0.7785 & 0.7377 & 0.8618 & -0.7463 & 0.3647 & 0.002082\\
M027 & 0.1229 & 0.0234 & 0.8976 & 0.8222 & 0.9698 & -1.0853 & 0.8683 & 0.002902\\
M028 & 0.1229 & 0.0231 & 0.8257 & 0.6109 & 0.9885 & -0.9311 & 0.8693 & 0.009086\\
M029 & 0.1274 & 0.0228 & 0.8999 & 0.8259 & 0.8505 & -0.7805 & 0.5688 & 0.006588\\
M030 & 0.1404 & 0.0222 & 0.8232 & 0.6852 & 0.8679 & -0.8594 & -0.4637 & 0.008126\\
M031 & 0.1386 & 0.0229 & 0.7693 & 0.6684 & 1.0478 & -1.2670 & 1.2536 & 0.006502\\
M032 & 0.1369 & 0.0215 & 0.8812 & 0.8019 & 1.0005 & -0.7282 & -1.6927 & 0.000905\\
M033 & 0.1286 & 0.0230 & 0.7005 & 0.6752 & 1.0492 & -0.7119 & -0.8184 & 0.007968\\
M034 & 0.1354 & 0.0216 & 0.7018 & 0.5970 & 0.8791 & -0.8252 & -1.1148 & 0.003620\\
M035 & 0.1359 & 0.0228 & 0.8210 & 0.6815 & 0.9872 & -1.1642 & -0.1801 & 0.004440\\
M036 & 0.1390 & 0.0220 & 0.8631 & 0.6477 & 0.8985 & -0.8632 & 0.8285 & 0.001082\\
\hline\hline
\end{tabular}
\end{center}
\end{table*}

\begin{table*}
\begin{center}
\caption{Second Set of Models from the Nested Design\label{app:tab3}}
\begin{tabular}{lcccccccc}
Model & $\omega_m$  & $\omega_b$ & $\sigma_8$ & $h$  & $n_s$ & $w_0$ & $w_a$ & $\omega_\nu$ \\
\hline\hline
M037 & 0.1539 & 0.0224 & 0.8529 & 0.5965 & 0.8943 & -1.2542 & 0.8868 & 0.003549\\
M038 & 0.1467 & 0.0227 & 0.7325 & 0.5902 & 0.9562 & -0.8019 & 0.3628 & 0.007077\\
M039 & 0.1209 & 0.0223 & 0.8311 & 0.7327 & 0.9914 & -0.7731 & 0.4896 & 0.001973\\
M040 & 0.1466 & 0.0229 & 0.8044 & 0.8015 & 0.9376 & -0.9561 & -0.0359 & 0.000893\\
M041 & 0.1274 & 0.0218 & 0.7386 & 0.6752 & 0.9707 & -1.2903 & 1.0416 & 0.003045\\
M042 & 0.1244 & 0.0230 & 0.7731 & 0.6159 & 0.8588 & -0.9043 & 0.8095 & 0.009194\\
M043 & 0.1508 & 0.0233 & 0.7130 & 0.8259 & 0.9676 & -1.0551 & 0.3926 & 0.009998\\
M044 & 0.1389 & 0.0224 & 0.8758 & 0.6801 & 0.9976 & -0.8861 & -0.1804 & 0.008018\\
M045 & 0.1401 & 0.0228 & 0.7167 & 0.6734 & 0.9182 & -1.2402 & 1.2155 & 0.006610\\
M046 & 0.1381 & 0.0224 & 0.7349 & 0.8277 & 1.0202 & -1.1052 & -1.0533 & 0.006433\\
M047 & 0.1411 & 0.0216 & 0.7770 & 0.7939 & 0.9315 & -0.8042 & 0.7010 & 0.003075\\
M048 & 0.1374 & 0.0226 & 0.7683 & 0.6865 & 0.8576 & -1.1374 & -0.5106 & 0.004548\\
M049 & 0.1339 & 0.0217 & 0.7544 & 0.5920 & 1.0088 & -0.8520 & -0.7438 & 0.003512\\
M050 & 0.1337 & 0.0233 & 0.8092 & 0.7309 & 0.9389 & -0.7230 & 0.6920 & 0.005539\\
M051 & 0.1514 & 0.0222 & 0.7433 & 0.6502 & 0.8922 & -0.9871 & 0.8803 & 0.002842\\
M052 & 0.1483 & 0.0230 & 0.7012 & 0.6840 & 0.9809 & -1.2881 & -0.9045 & 0.006199\\
M053 & 0.1226 & 0.0226 & 0.7998 & 0.8265 & 1.0161 & -1.2593 & -0.3858 & 0.001096\\
M054 & 0.1345 & 0.0216 & 0.8505 & 0.6251 & 0.8535 & -1.2526 & 0.5703 & 0.007438\\
M055 & 0.1298 & 0.0222 & 0.7504 & 0.8170 & 0.9574 & -1.0573 & 1.0338 & 0.006843\\
M056 & 0.1529 & 0.0219 & 0.8508 & 0.6438 & 1.0322 & -0.7359 & 0.6931 & 0.006311\\
M057 & 0.1419 & 0.0234 & 0.7937 & 0.7415 & 1.0016 & -0.7710 & -1.5964 & 0.005128\\
M058 & 0.1226 & 0.0224 & 0.7278 & 0.6152 & 1.0348 & -1.1051 & 0.2955 & 0.007280\\
M059 & 0.1529 & 0.0224 & 0.7035 & 0.6877 & 0.8616 & -0.9833 & -1.1788 & 0.009885\\
M060 & 0.1270 & 0.0233 & 0.8827 & 0.5622 & 0.8609 & -1.1714 & 0.8346 & 0.009901\\
M061 & 0.1272 & 0.0220 & 0.8021 & 0.8302 & 0.8968 & -0.9545 & -0.6659 & 0.004782\\
M062 & 0.1257 & 0.0216 & 0.7699 & 0.5813 & 0.9460 & -0.8041 & 0.7956 & 0.003922\\
M063 & 0.1312 & 0.0229 & 0.7974 & 0.5890 & 0.9522 & -0.9560 & 0.6650 & 0.001740\\
M064 & 0.1437 & 0.0218 & 0.8866 & 0.7402 & 0.9335 & -1.0713 & 0.7128 & 0.003782\\
M065 & 0.1445 & 0.0218 & 0.8797 & 0.5554 & 0.9353 & -1.2423 & -1.3032 & 0.008850\\
\hline\hline
\end{tabular}
\end{center}
\end{table*}

\begin{table*}
\begin{center}
\caption{Third Set of Models from the Nested Design\label{app:tab4}}
\begin{tabular}{lcccccccc}
Model & $\omega_m$  & $\omega_b$ & $\sigma_8$ & $h$  & $n_s$ & $w_0$ & $w_a$ & $\omega_\nu$ \\
\hline\hline
M066 & 0.1392 & 0.0229 & 0.8849 & 0.8176 & 0.8579 & -1.2334 & 0.7098 & 0.000187\\
M067 & 0.1398 & 0.0224 & 0.7795 & 0.7473 & 1.0392 & -1.0471 & 0.7377 & 0.008256\\
M068 & 0.1319 & 0.0232 & 0.7645 & 0.8112 & 0.9199 & -0.7812 & 0.1646 & 0.003716\\
M069 & 0.1392 & 0.0227 & 0.8130 & 0.6062 & 0.8765 & -1.0792 & 0.8817 & 0.006372\\
M070 & 0.1499 & 0.0232 & 0.8093 & 0.7587 & 0.9260 & -0.8942 & -0.9967 & 0.009760\\
M071 & 0.1218 & 0.0224 & 0.7348 & 0.7999 & 1.0330 & -0.9341 & 0.8896 & 0.002212\\
M072 & 0.1265 & 0.0218 & 0.8349 & 0.6080 & 0.9291 & -1.1293 & 0.2197 & 0.002806\\
M073 & 0.1212 & 0.0227 & 0.7683 & 0.6587 & 0.8704 & -0.9662 & 0.9453 & 0.000327\\
M074 & 0.1337 & 0.0216 & 0.8575 & 0.8099 & 0.8518 & -1.0815 & 1.0506 & 0.002370\\
M075 & 0.1484 & 0.0230 & 0.8491 & 0.7212 & 0.9566 & -0.8980 & 0.8181 & 0.002716\\
M076 & 0.1419 & 0.0215 & 0.7700 & 0.6091 & 0.9332 & -0.9753 & -0.2691 & 0.008143\\
M077 & 0.1321 & 0.0230 & 0.7011 & 0.6562 & 0.9938 & -1.1170 & 1.0706 & 0.001979\\
M078 & 0.1446 & 0.0219 & 0.7903 & 0.8074 & 0.9752 & -1.2323 & 1.1681 & 0.004021\\
M079 & 0.1344 & 0.0235 & 0.8742 & 0.7575 & 0.9219 & -1.0483 & -0.3793 & 0.004423\\
M080 & 0.1419 & 0.0218 & 0.8420 & 0.8205 & 0.9145 & -1.1295 & -1.2357 & 0.001959\\
M081 & 0.1366 & 0.0224 & 0.7034 & 0.6599 & 0.8745 & -0.8122 & 0.7675 & 0.005665\\
M082 & 0.1511 & 0.0218 & 0.8061 & 0.7242 & 1.0132 & -0.7940 & 0.0740 & 0.004488\\
M083 & 0.1501 & 0.0230 & 0.7458 & 0.6037 & 0.9999 & -1.2300 & 0.9126 & 0.008023\\
M084 & 0.1244 & 0.0227 & 0.8444 & 0.7462 & 1.0351 & -1.2012 & 1.0042 & 0.002919\\
M085 & 0.1546 & 0.0227 & 0.8201 & 0.8187 & 0.8620 & -1.0794 & 0.3306 & 0.005524\\
M086 & 0.1289 & 0.0221 & 0.8467 & 0.7498 & 0.9158 & -0.8964 & 0.7044 & 0.006605\\
M087 & 0.1437 & 0.0235 & 0.8383 & 0.6612 & 1.0206 & -0.7128 & 0.3537 & 0.006952\\
M088 & 0.1237 & 0.0229 & 0.8779 & 0.6050 & 0.8725 & -1.2333 & 1.1145 & 0.001034\\
M089 & 0.1505 & 0.0221 & 0.8396 & 0.5830 & 0.8506 & -0.8262 & 0.3234 & 0.002603\\
M090 & 0.1439 & 0.0226 & 0.8366 & 0.6987 & 0.9116 & -1.2874 & 0.2509 & 0.009071\\
M091 & 0.1325 & 0.0224 & 0.8076 & 0.6680 & 0.9435 & -0.7361 & -0.9543 & 0.003494\\
M092 & 0.1476 & 0.0215 & 0.8452 & 0.8060 & 0.9855 & -0.9543 & 0.9158 & 0.007598\\
M093 & 0.1389 & 0.0219 & 0.7151 & 0.6105 & 0.9228 & -1.2533 & -0.3018 & 0.004566\\
M094 & 0.1386 & 0.0235 & 0.7699 & 0.7494 & 0.8529 & -1.1243 & 1.0953 & 0.006593\\
M095 & 0.1424 & 0.0223 & 0.8764 & 0.6612 & 0.9422 & -1.2912 & 1.2729 & 0.002028\\
M096 & 0.1404 & 0.0219 & 0.8946 & 0.8155 & 1.0442 & -1.1562 & -0.8081 & 0.001851\\
M097 & 0.1351 & 0.0226 & 0.7560 & 0.6549 & 1.0041 & -0.8389 & 0.8124 & 0.005556\\
M098 & 0.1259 & 0.0225 & 0.7918 & 0.7512 & 0.9055 & -1.1744 & 0.9019 & 0.003027\\
M099 & 0.1237 & 0.0233 & 0.8907 & 0.6375 & 0.9715 & -1.2563 & -0.6293 & 0.007970\\
M100 & 0.1232 & 0.0221 & 0.7715 & 0.5530 & 0.8635 & -0.9174 & -1.7275 & 0.007150\\
M101 & 0.1526 & 0.0217 & 0.7535 & 0.7292 & 0.8836 & -0.7672 & -0.1455 & 0.004596\\
M102 & 0.1531 & 0.0229 & 0.8727 & 0.8137 & 0.9916 & -1.1062 & 0.5227 & 0.005416\\
M103 & 0.1274 & 0.0223 & 0.8993 & 0.7448 & 1.0455 & -0.9231 & 0.7886 & 0.006497\\
M104 & 0.1481 & 0.0222 & 0.7512 & 0.7255 & 1.0029 & -1.0720 & 0.1433 & 0.000910\\
M105 & 0.1490 & 0.0222 & 0.8922 & 0.5780 & 0.9802 & -0.8530 & 0.4716 & 0.002495\\
M106 & 0.1422 & 0.0223 & 0.8679 & 0.6048 & 0.8869 & -0.8012 & 0.6662 & 0.009949\\
M107 & 0.1304 & 0.0233 & 0.8171 & 0.8062 & 1.0495 & -0.8080 & 0.3337 & 0.003607\\
M108 & 0.1414 & 0.0223 & 0.7269 & 0.7524 & 0.9095 & -1.0203 & 0.6065 & 0.008364\\
M109 & 0.1377 & 0.0229 & 0.8656 & 0.6012 & 1.0062 & -1.1060 & 0.9672 & 0.006263\\
M110 & 0.1336 & 0.0220 & 0.8703 & 0.8423 & 0.9509 & -1.1045 & 0.0875 & 0.009305\\
M111 & 0.1212 & 0.0230 & 0.7810 & 0.6912 & 0.9695 & -0.9892 & 0.1224 & 0.007263\\
\hline\hline
\end{tabular}
\end{center}
\end{table*}

\end{appendix}



\clearpage

\pagebreak
\pagestyle{plain}
\onecolumn
\begin{center}
\textbf{\LARGE Supplemental Materials: ``The Mira-Titan Universe IV. High Precision Power Spectrum Emulation''}
\vspace{0.25in}

{\Large By Kelly R. Moran, Katrin Heitmann, Earl Lawrence, Salman Habib, Derek Bingham, Amol Upadhye, Juliana Kwan, David Higdon, Richard Payne}

\vspace{0.25in}
\end{center}
\setcounter{equation}{0}
\setcounter{figure}{0}
\setcounter{table}{0}
\setcounter{page}{1}
\setcounter{section}{0}
\makeatletter
\renewcommand{\thesection}{S\arabic{section}}
\renewcommand{\theequation}{S\arabic{equation}}
\renewcommand{\thefigure}{S\arabic{figure}}
\renewcommand{\bibnumfmt}[1]{[S#1]}
\renewcommand{\citenumfont}[1]{S#1}
\renewcommand{\thepage}{S\arabic{page}}

The supplemental material provides additional visualizations and metrics for the data alignment step, the process convolution smoothing, and the $P_{cb}$ and $P_{tot}$ emulator performance.

\section{Data Alignment}

\begin{figure*}
\centering
\includegraphics[width=0.9\textwidth]{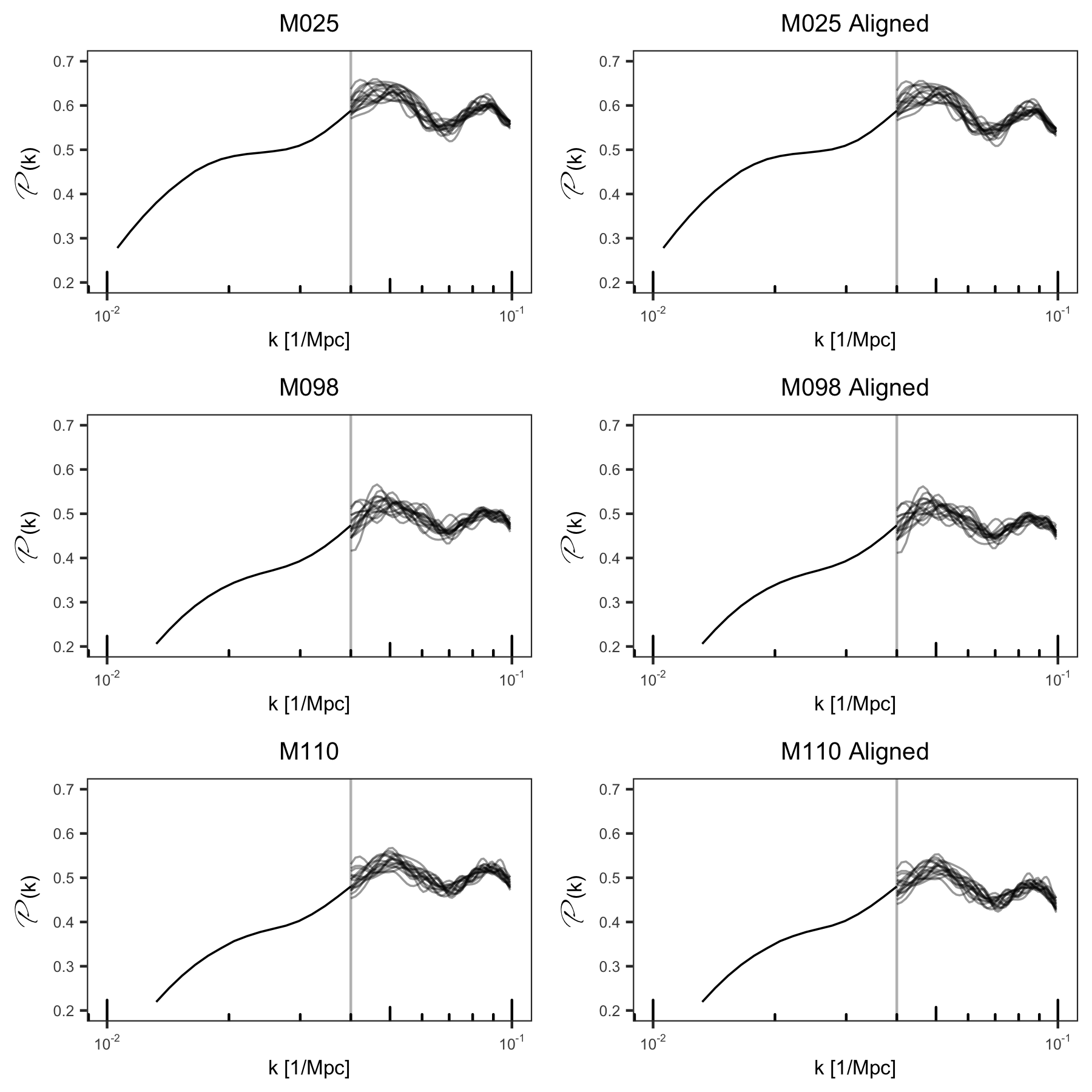}
\caption{PT and LR realizations for two example cosmologies before (left column) and after (right column) alignment of the LR simulations to the PT about the match point. PT realizations are to the left of the vertical line in each subplot and LR simulation realizations are right of the vertical line in each subplot.\label{fig:supplement_alignment}}
\end{figure*}

\begin{figure*}
\centering
\includegraphics[width=0.9\textwidth]{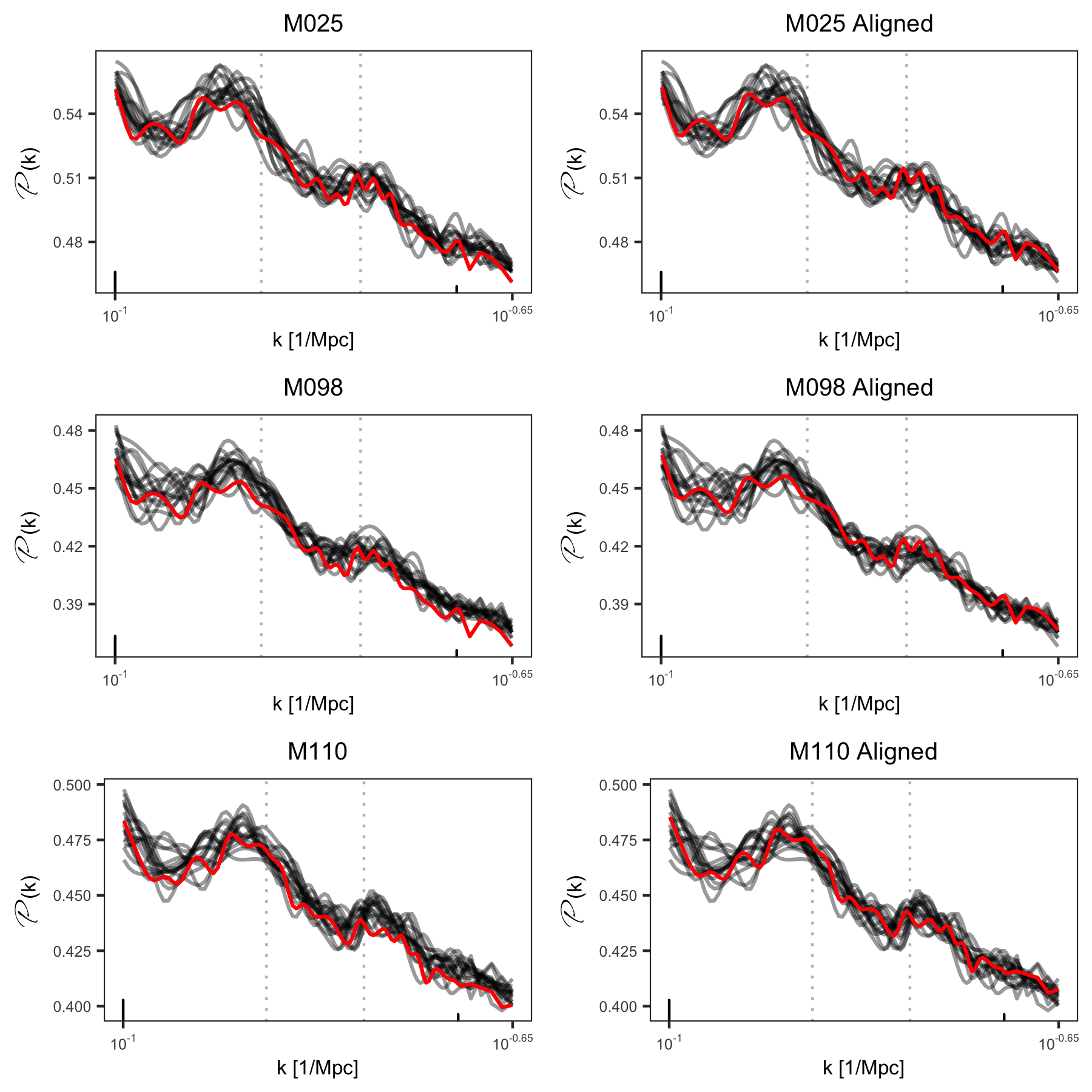}
\caption{LR and HR simulations for two example cosmologies before (left column) and after (right column) alignment of the HR simulations to the LR within a match window. LR simulations are the black lines and the HR simulation is the red lines in each subplot, while the match window is shown via a dashed grey line.\label{fig:supplement_alignment_2}}
\end{figure*}

Examples of transformed spectra (on the emulation scale) near the match windows before and after data alignment are shown in Figures \ref{fig:supplement_alignment} and \ref{fig:supplement_alignment_2}. In both plots, the redshift is $z=2.02$. Alignment allows for more sensible smoothing via two-layer process convolution around the matching point.

\section{Process Convolution Diagnostics}

\begin{figure*}
\includegraphics[width=0.75\textwidth]{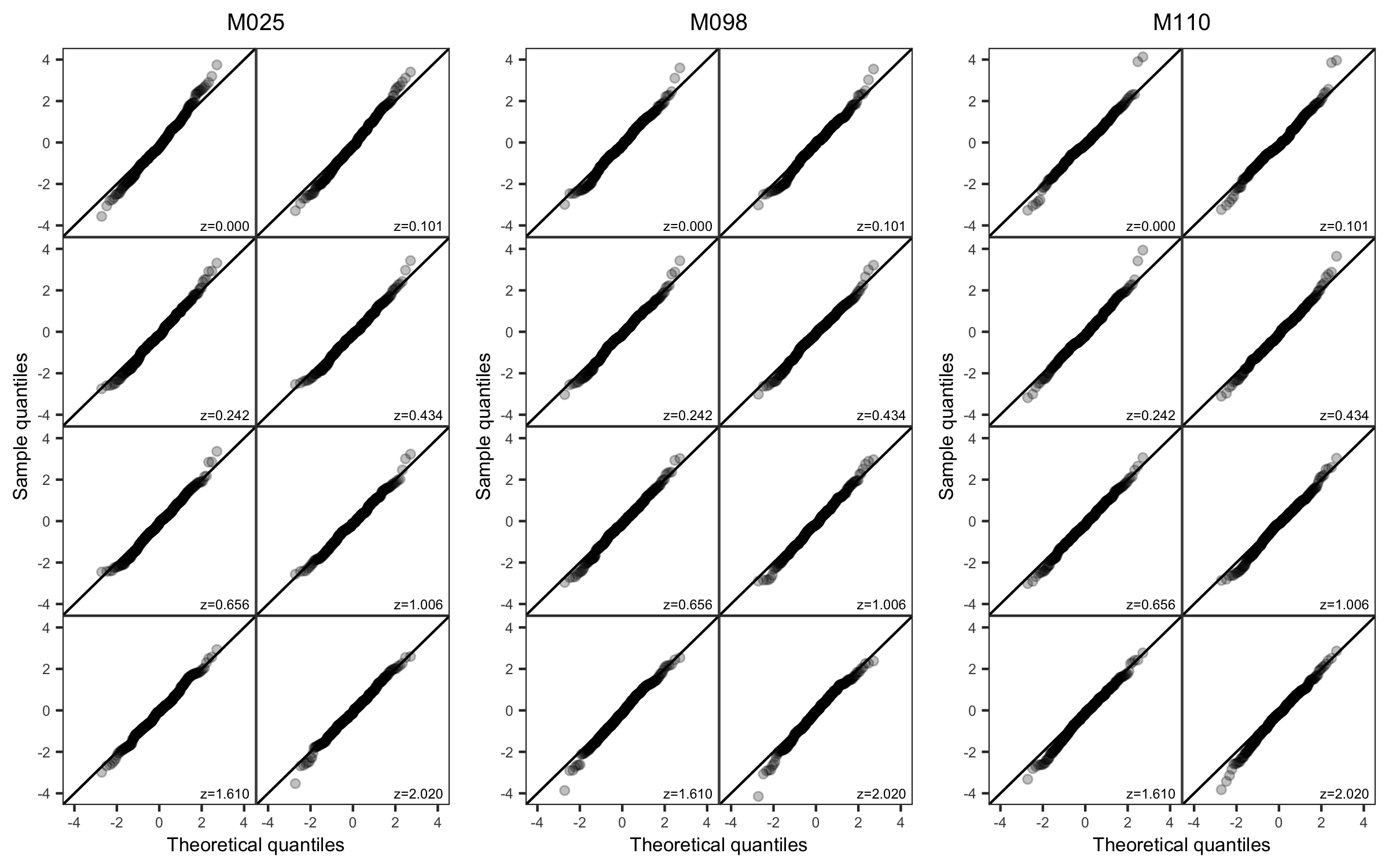}
\caption{Quantile-quantile plots of the standardized residuals for the high-resolution simulations at the eight redshifts $z$ for three cosmologies. Standardized residuals are computed by subtracting the estimated mean from the high-resolution simulation and multiplying each value by the square root of the precision at its $k$ value. We expect the resulting sample to follow a standard normal distribution with no dependence on $k$. The sample quantiles of the standardized residuals are plotted against the theoretical quantiles for a sample of the same size from the standard normal distribution. The tails diverging from the $x=y$ line indicate slightly heavier tails than expected from normality, but overall the model fits relatively well.}
\label{fig:diagnostics_qq_highres}
\end{figure*}

We provide quantile-quantile plots of these standardized residuals, broken down by HR-only (Figure \ref{fig:diagnostics_qq_highres}), LR-only (Figure \ref{fig:diagnostics_qq_lowres}), and combined (Figure \ref{fig:diagnostics_qq_bothres}) residuals. An approximate match to the $45^{\circ}$ line by the points indicates a good fit.

\begin{figure*}
\includegraphics[width=0.7\textwidth]{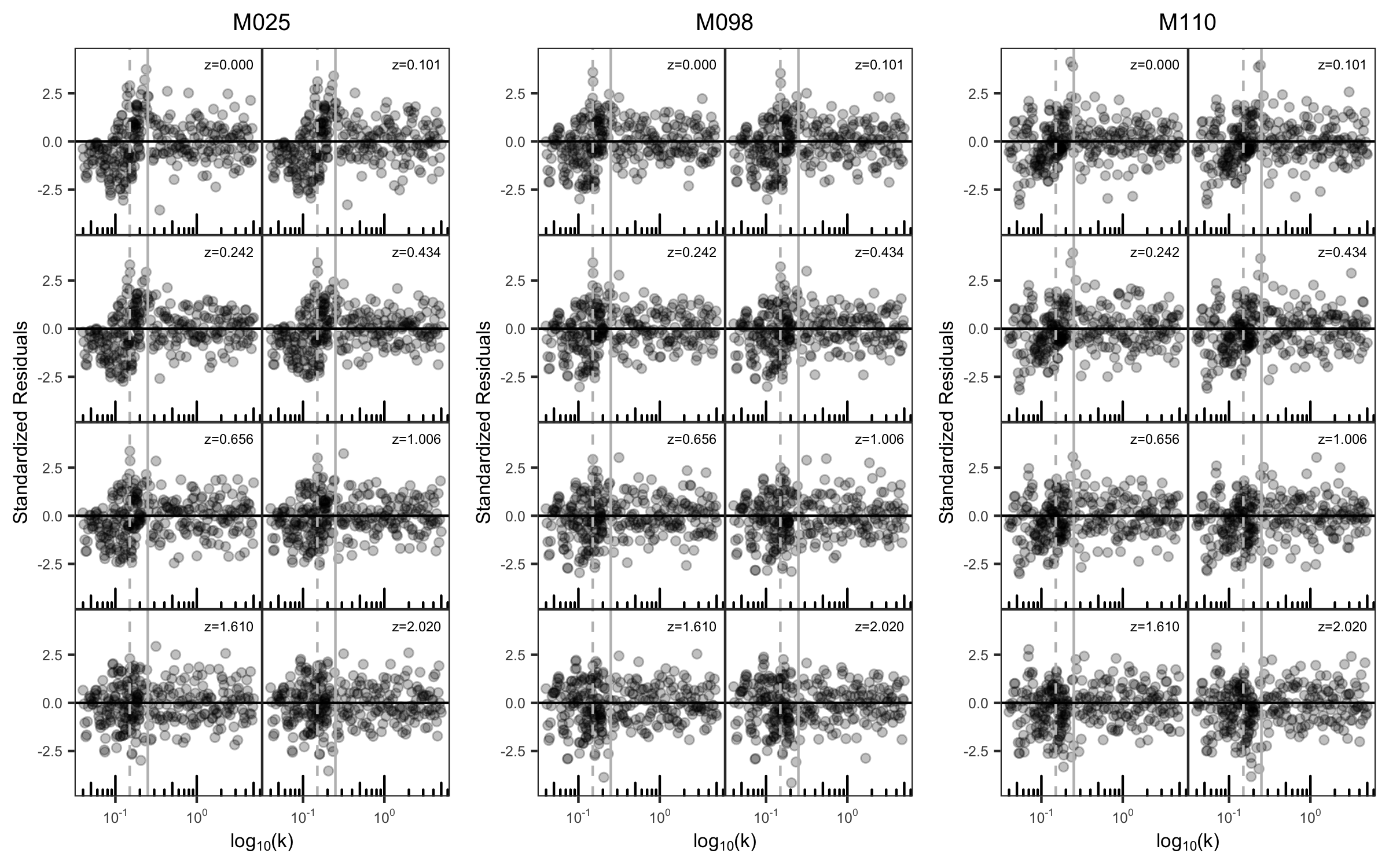}
\caption{Standardized residuals for the high-resolution simulation at the eight redshifts $z$ for three cosmologies plotted against $k$. There is a slight trend present for some cosmology and redshift combinations in moving from low $k$ to high $k$ towards the point where the low res simulations are no longer included in the smoothing. Other than this feature, no obvious correlations or structure are visible. The dashed vertical line is the midpoint of the window at which the HR simulations are aligned to the LR simulations, and the solid vertical line is the point at which LR simulations are no longer included in the smoothing.}
\label{fig:diagnostics_resid_highres}
\end{figure*}

\begin{figure*}
\includegraphics[width=0.7\textwidth]{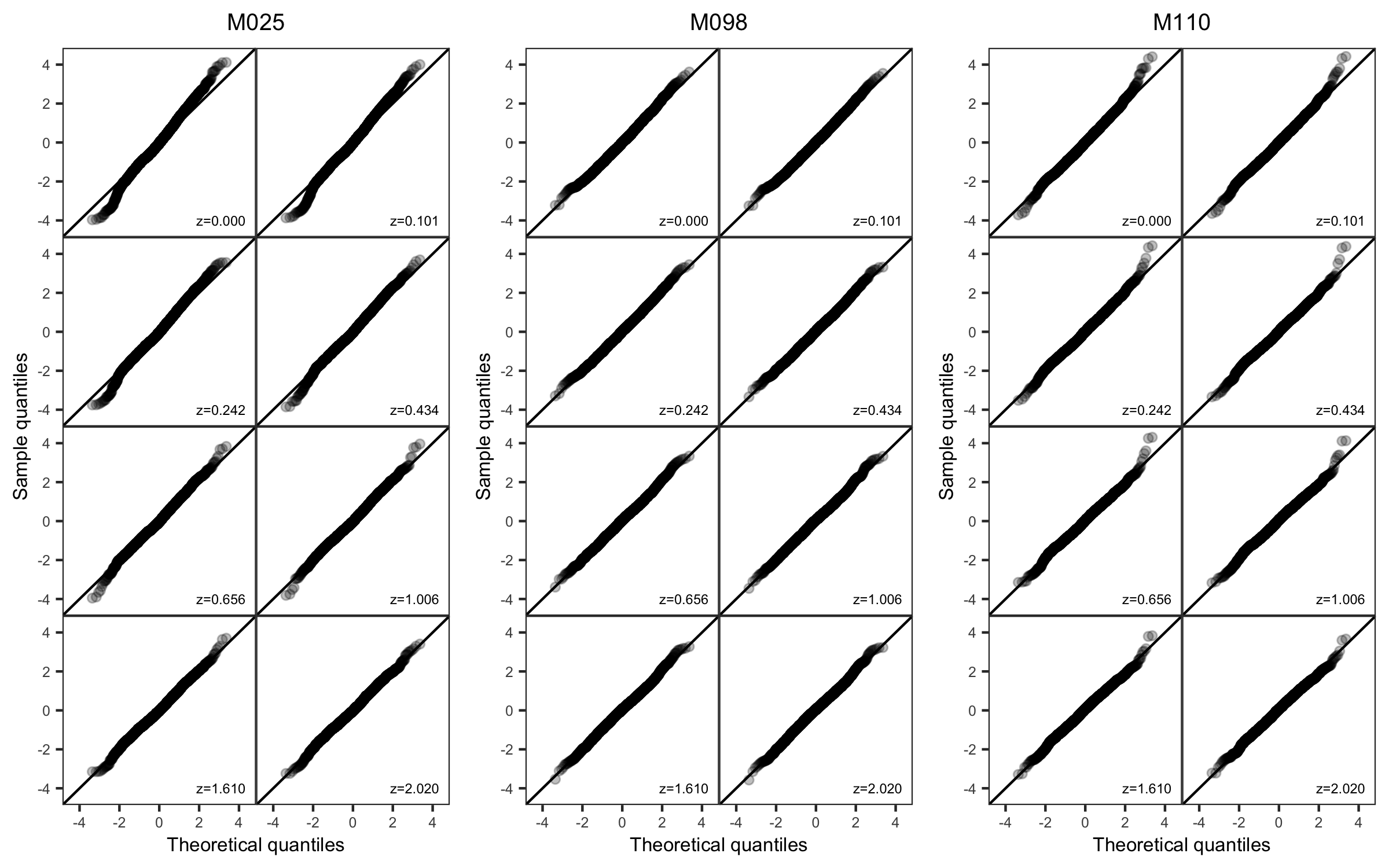}
\caption{Quantile-quantile plots of the standardized residuals for the lower-resolution simulations at the eight redshifts $z$ for three cosmologies.}
\label{fig:diagnostics_qq_lowres}
\end{figure*}

\begin{figure*}
\includegraphics[width=0.7\textwidth]{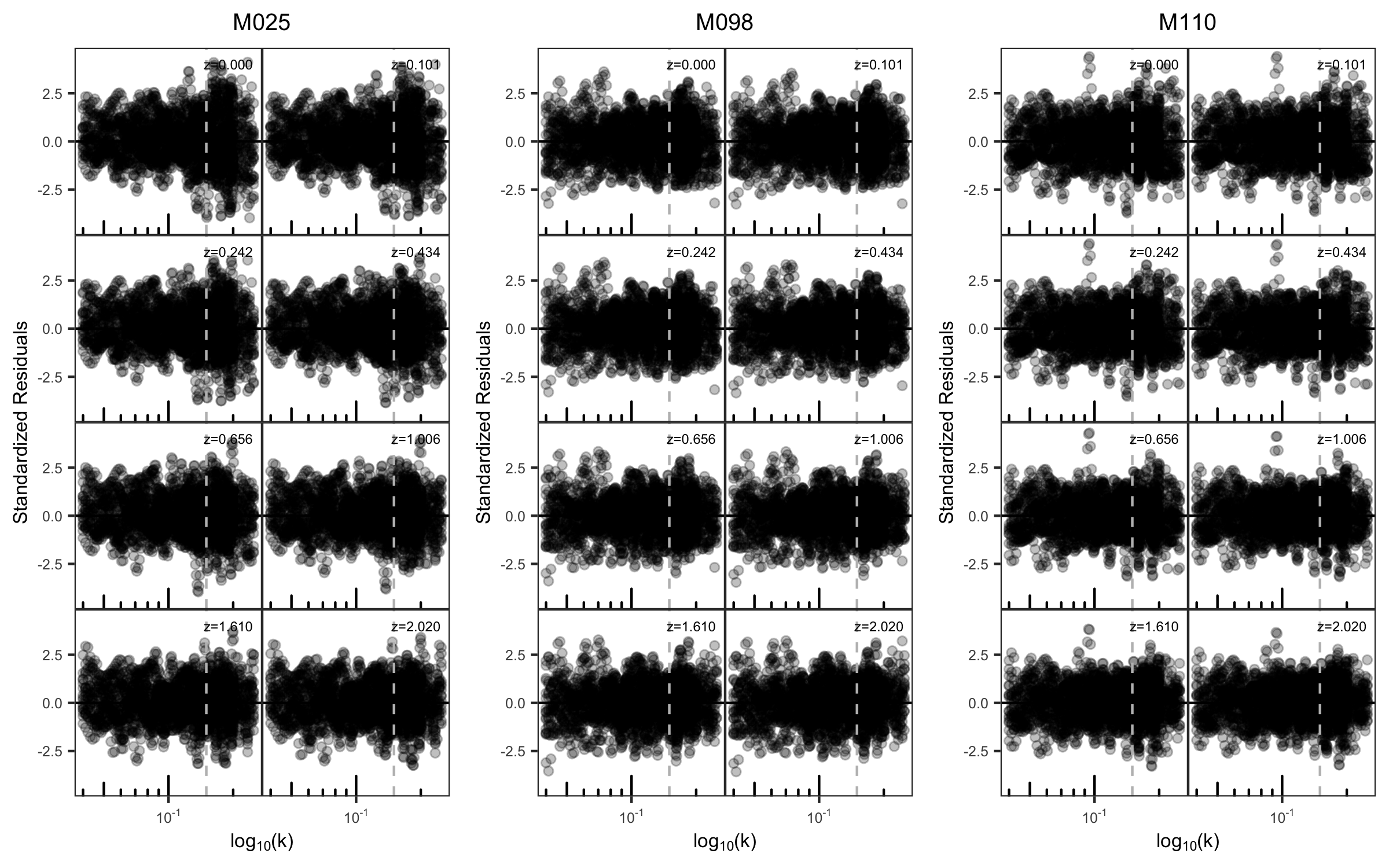}
\caption{Standardized residuals for the lower-resolution simulations at the eight redshifts $z$ for three cosmologies plotted against $k$.}
\label{fig:diagnostics_resid_lowres}
\end{figure*}

\begin{figure*}
\includegraphics[width=0.7\textwidth]{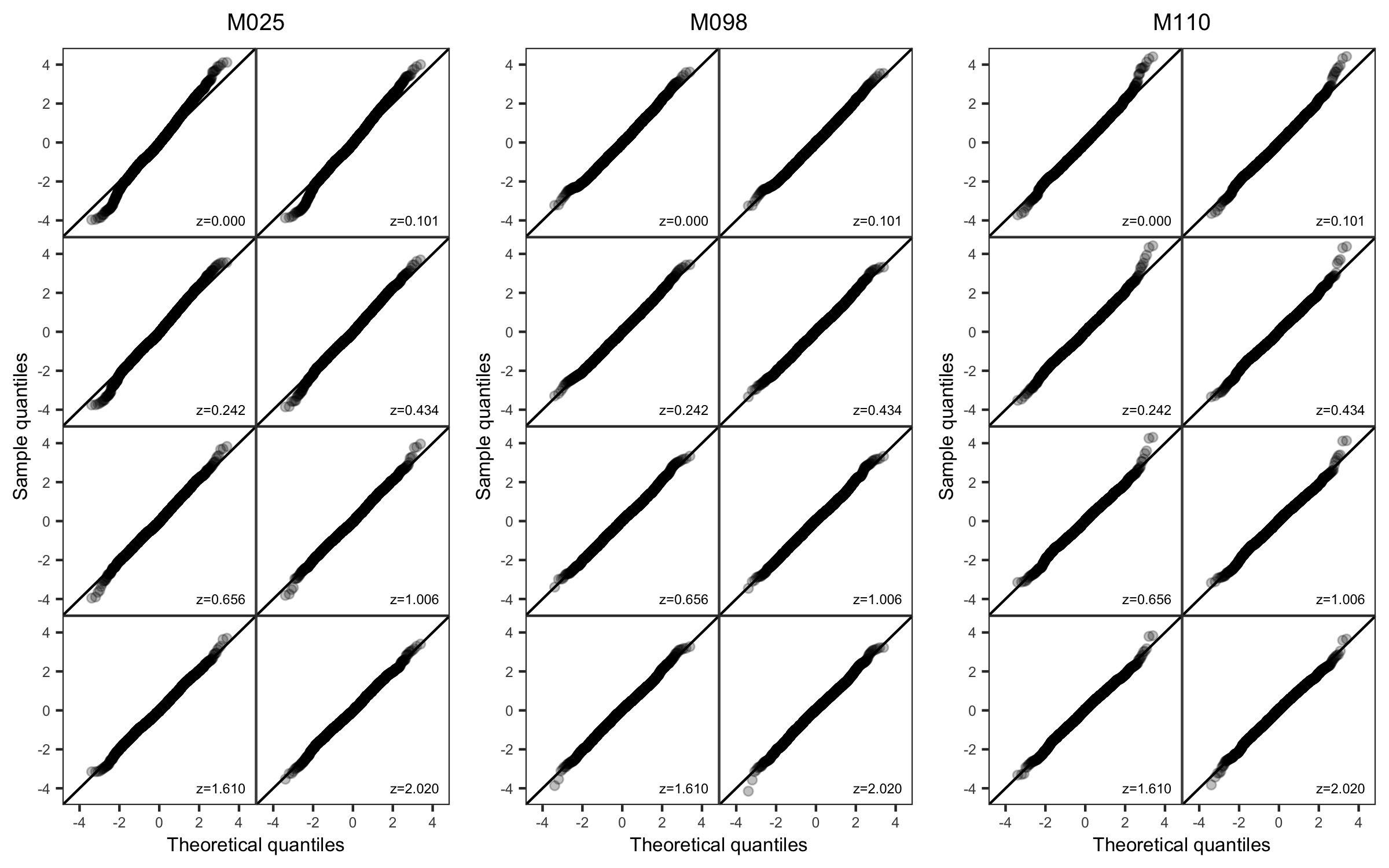}
\caption{Quantile-quantile plots of the standardized residuals for the lower-resolution and higher-resolution simulations at the eight redshifts $z$ for three cosmologies.}
\label{fig:diagnostics_qq_bothres}
\end{figure*}

\begin{figure*}
\includegraphics[width=0.7\textwidth]{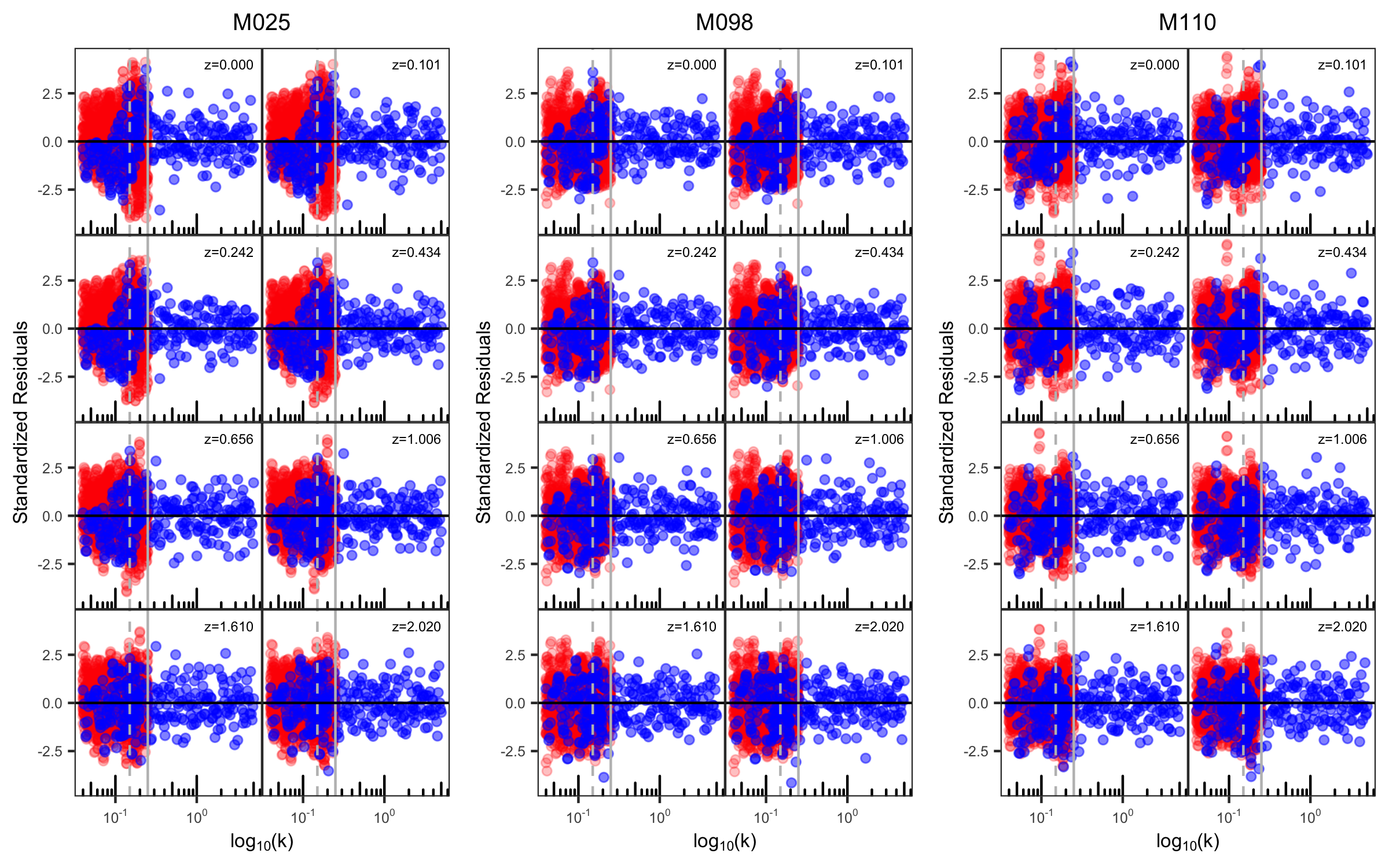}
\caption{Standardized residuals for both the low-resolution (red points) and the high-resolution (blue points) simulations at the eight redshifts $z$ for three cosmologies plotted against $k$.}
\label{fig:diagnostics_resid_bothres}
\end{figure*}

We also show plots of these standardized residuals against $k$ broken down by HR-only (Figure \ref{fig:diagnostics_resid_highres}), LR-only (Figure \ref{fig:diagnostics_resid_lowres}), and combined (Figure \ref{fig:diagnostics_resid_bothres}) residuals. The structure seen in the early $k$ ranges for the HR standardized residuals indicates some violation of the independence assumption -- this structure is likely due to the alignment process. Specifically, there is a visible positive trend from the start of the LR simulations to the LR cutoff point for some cosmologies that is particularly evident at low redshifts. It appears that the residuals are consistently zero-centered around the point at which the HR simulations are aligned to the LR simulations, and again when there are only HR simulations.

\section{Additional Emulator Performance Results}

\begin{figure}
\centering
\includegraphics[width=0.45\textwidth]{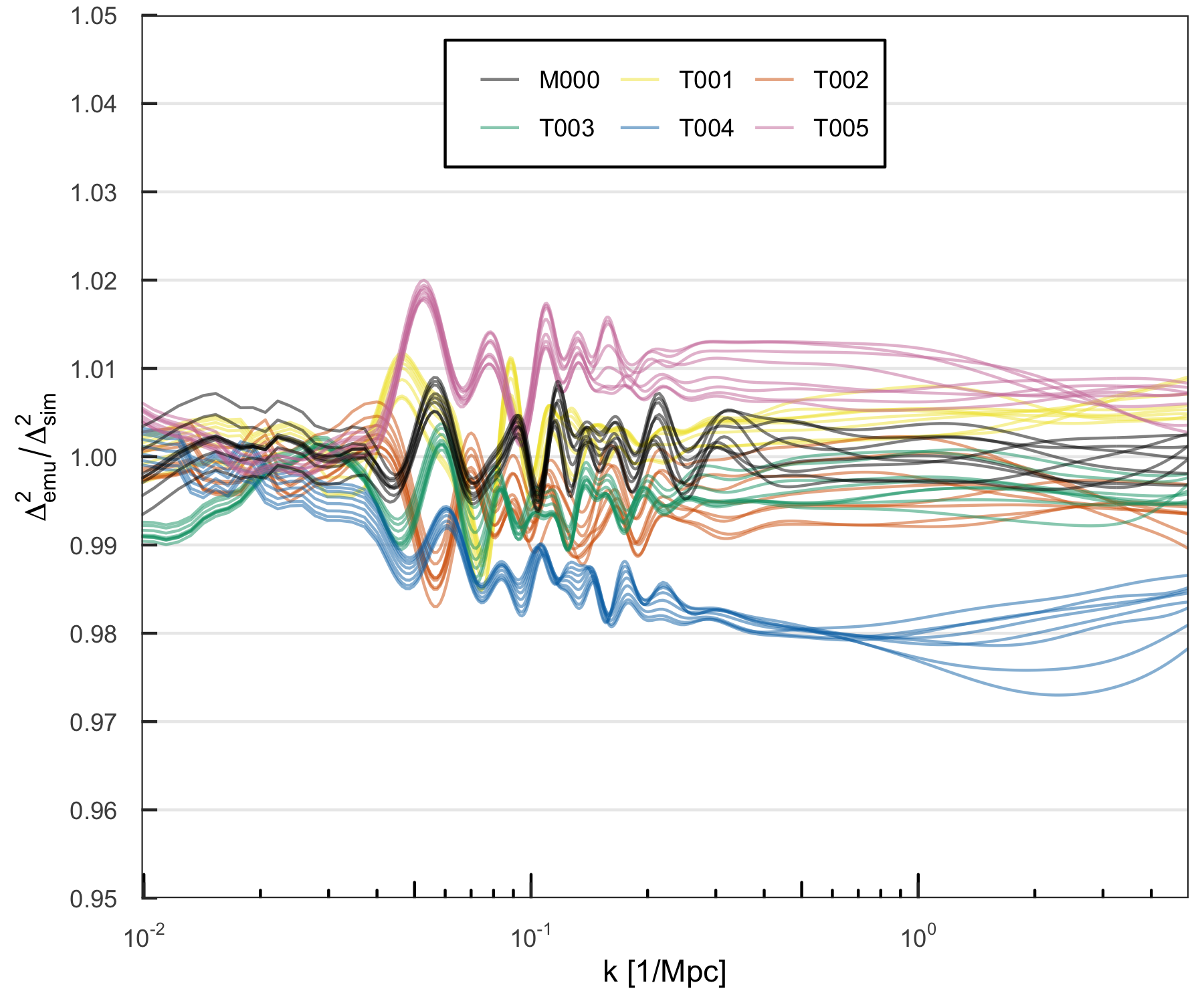}
\includegraphics[width=0.45\textwidth]{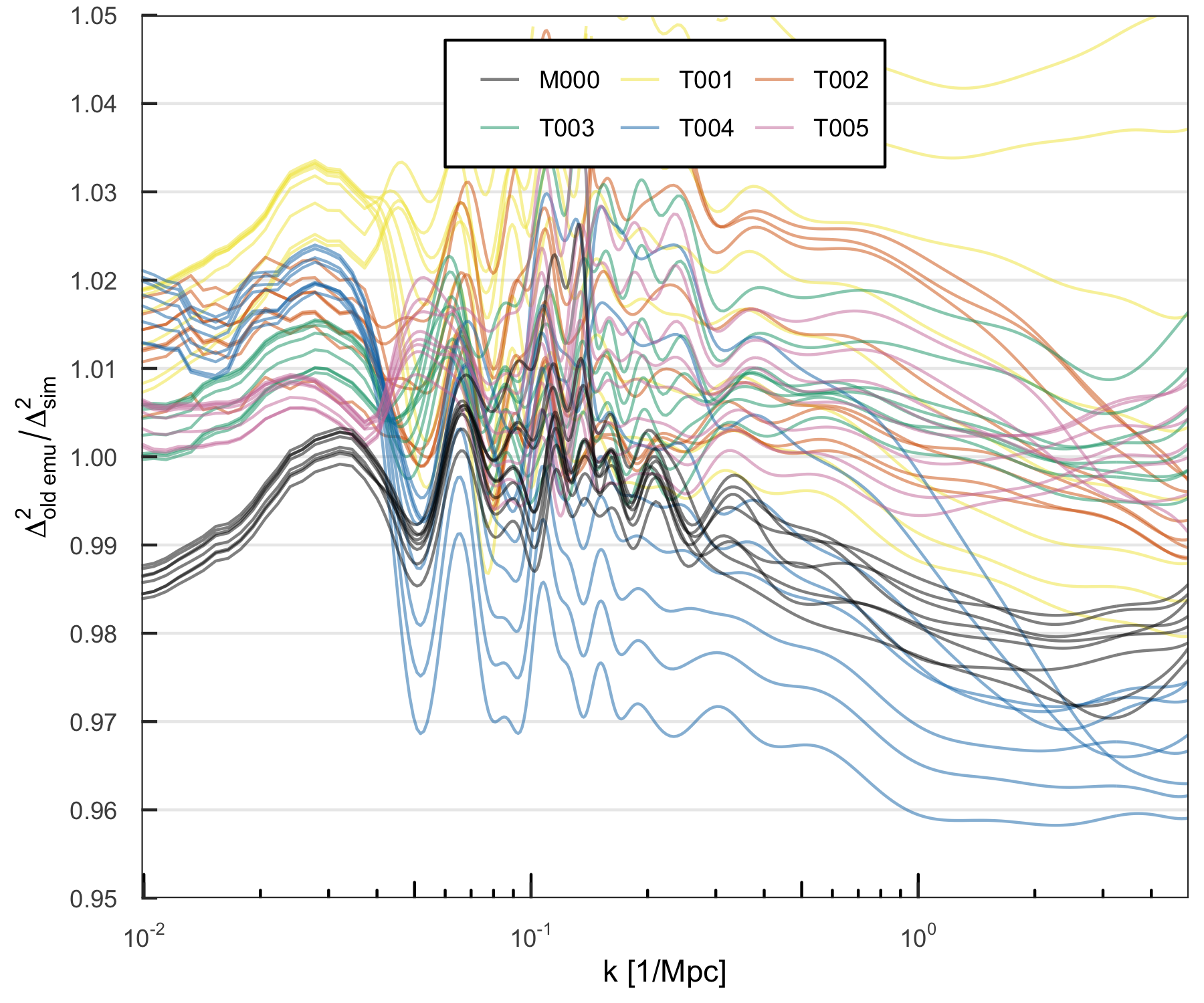}
\caption{Test of the final (left) and old (right) emulator accuracy for $P_{cb}$, the CDM and baryon component of the matter power spectrum. Different lines within one color are results for different redshifts. Upper panel: Results for the final Mira-Titan emulator based on 111 models. For all but T004, which has an absolute accuracy of 2.7\%, the accuracy is better than 2\%. Lower panel: For comparison, we show the emulator results from the first set of Mira-Titan simulations based on 36 cosmological models \citep{MT-pow1}. The original emulator is accurate at the 3-6\% level; specifically, the test runs shown have ratio $\Delta^2_{\text{emu}} / \Delta^2_{\text{sim}}$ ranging from 0.958 to 1.069.\label{fig:pcb}}
\end{figure}

\begin{figure}
\centering
\includegraphics[width=0.5\textwidth]{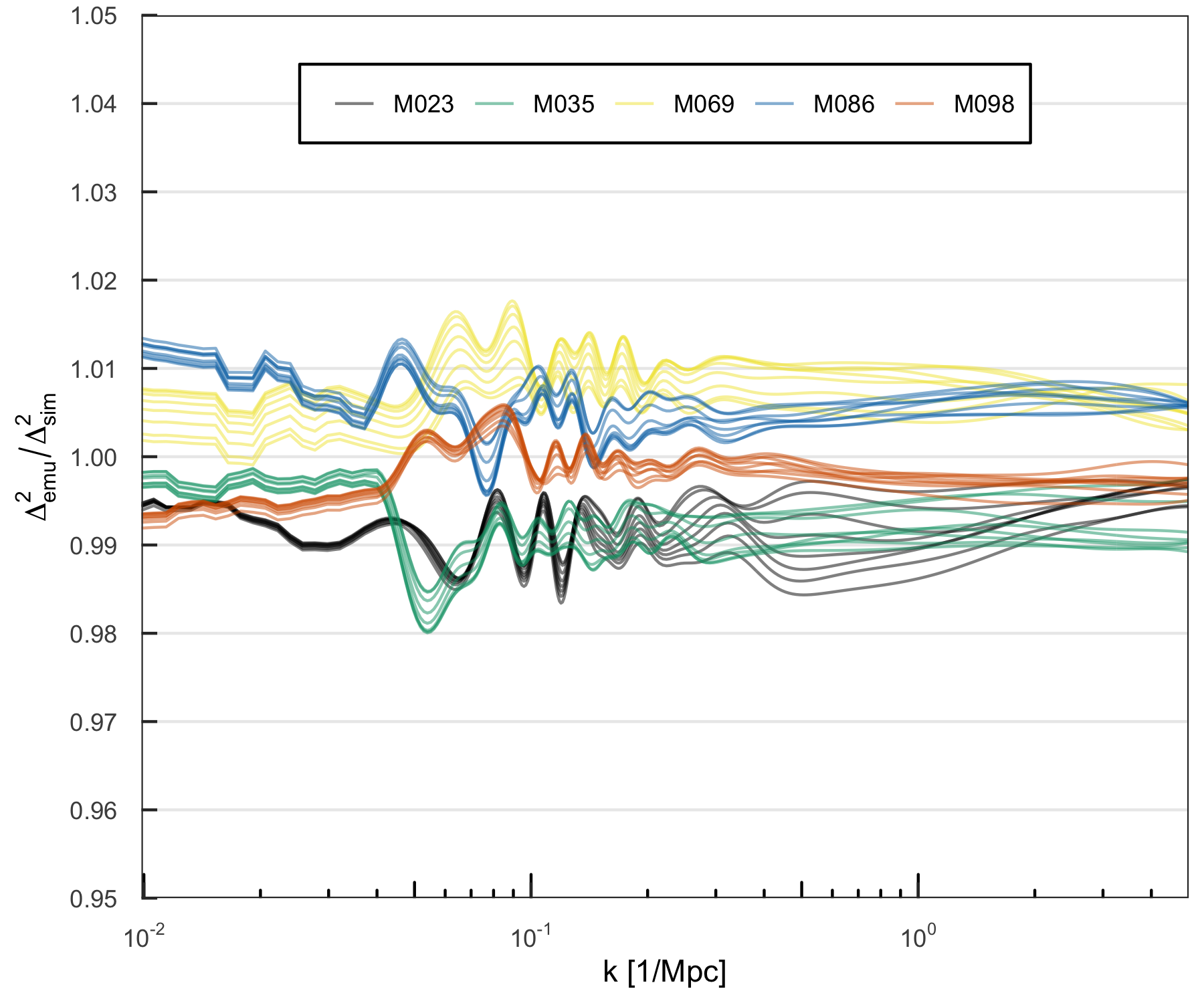}
\caption{Leave-out-one cross-validation results for the $P_{cb}$ emulator on the five cosmologies closest to the center of the design. Different lines within one color are results for different redshifts.\label{fig:perf_crossval_pcb}}
\end{figure}

\begin{figure}
\includegraphics[width=0.45\textwidth]{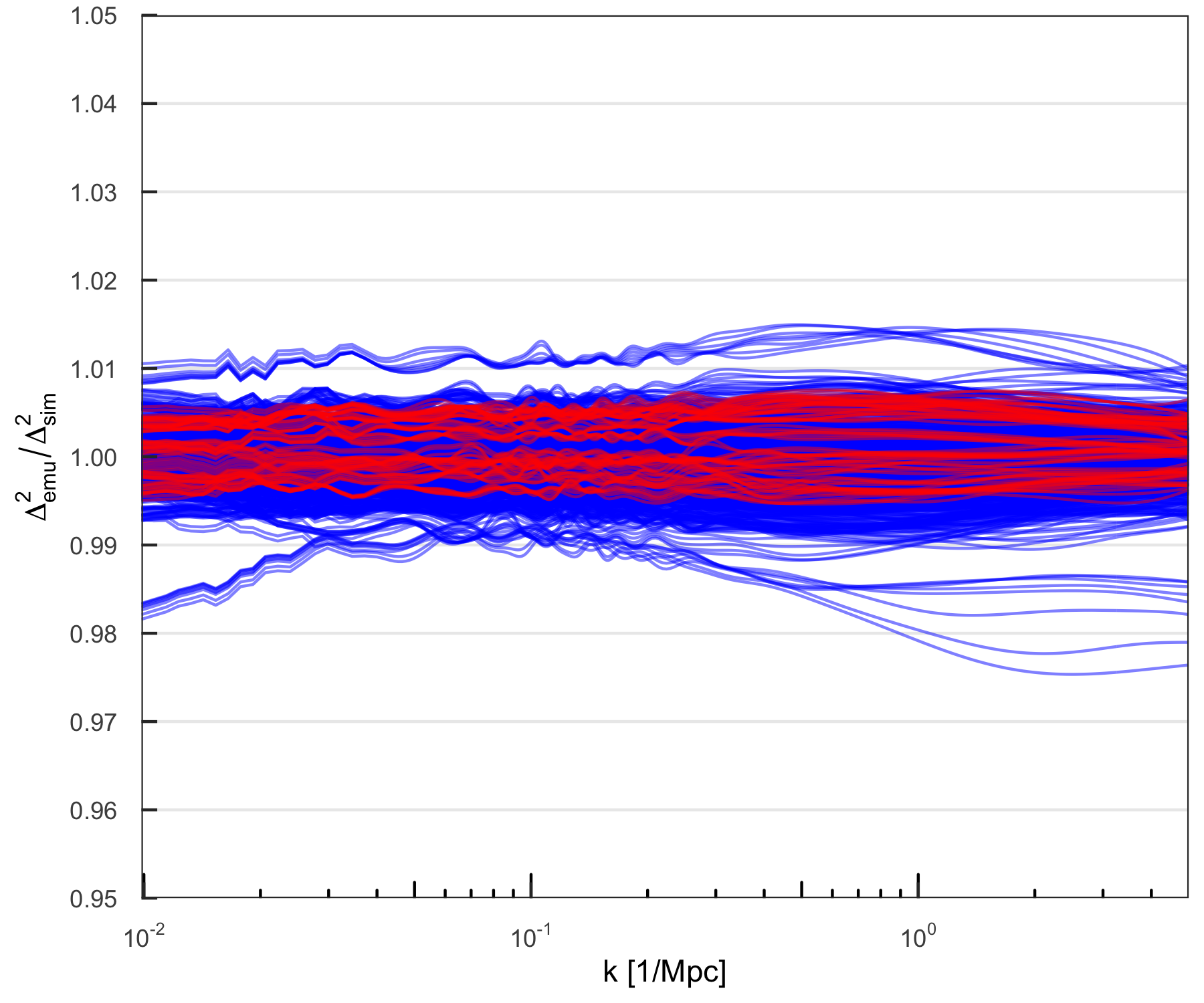}
\includegraphics[width=0.45\textwidth]{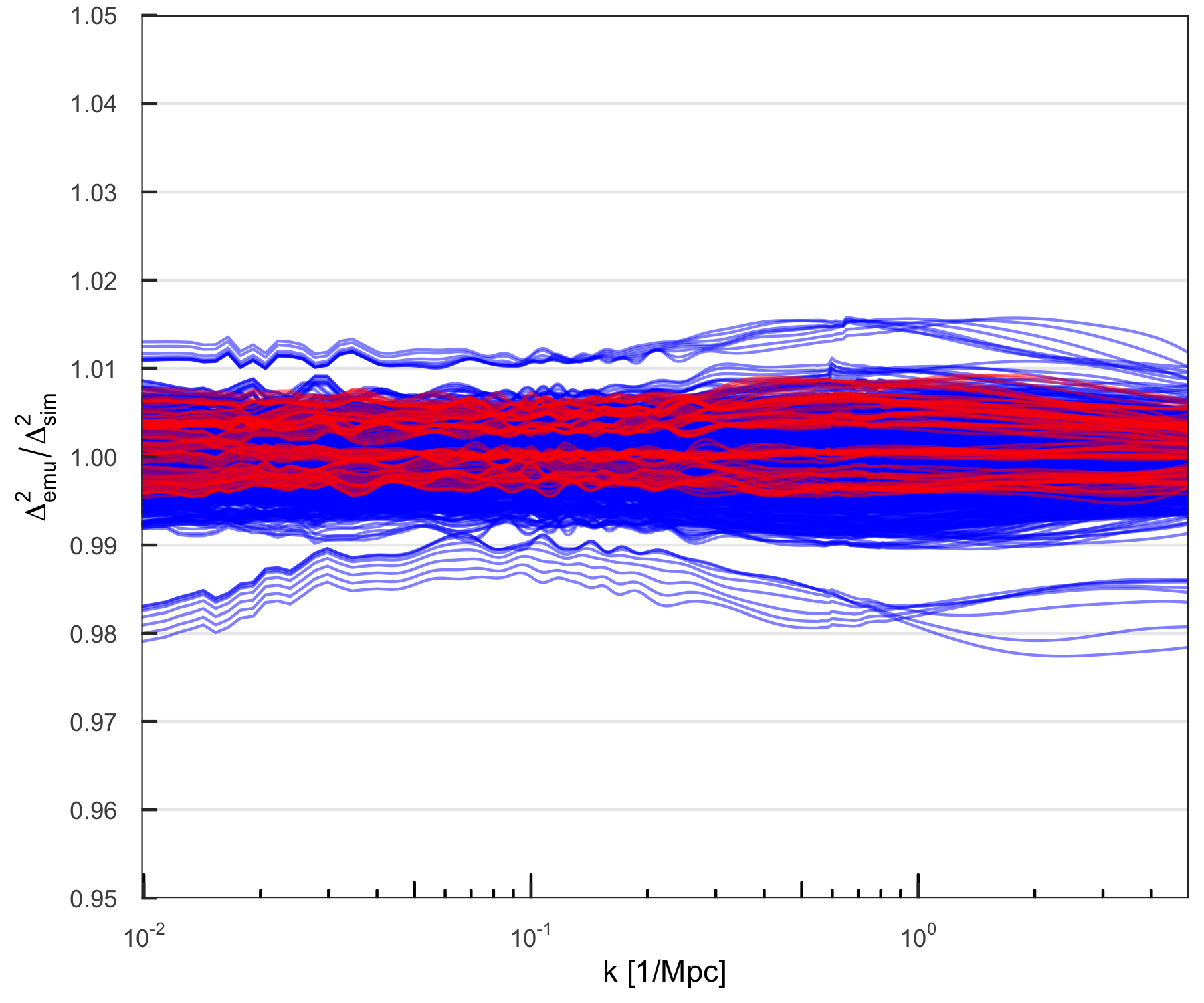}
\caption{In-sample predictions for for $P_{cb}$ (left) and $P_{tot}$ (right) emulators. The ratio $\Delta^2_{\text{emu}} / \Delta^2_{\text{sim}}$ ranges from 0.975 to 1.015 for $P_{cb}$, and from 0.977 to 1.016 for $P_{tot}$. Red (blue) lines correspond to cosmologies having (non)zero neutrino mass.\label{fig:perf_train}}
\end{figure}

Figure \ref{fig:pcb} shows tests of the emulator accuracy for $P_{cb}$, the CDM and baryon component of the matter power spectrum. Figure \ref{fig:perf_crossval_pcb} shows leave-out-one cross-validation results for the $P_{cb}$ emulators on the five cosmologies closest to the center of the design. In-sample predictions for $P_{cb}$ and $P_{tot}$ with the new emulator are shown in Figure \ref{fig:perf_train}.

\begin{figure}
\centering
\includegraphics[width=0.5\textwidth]{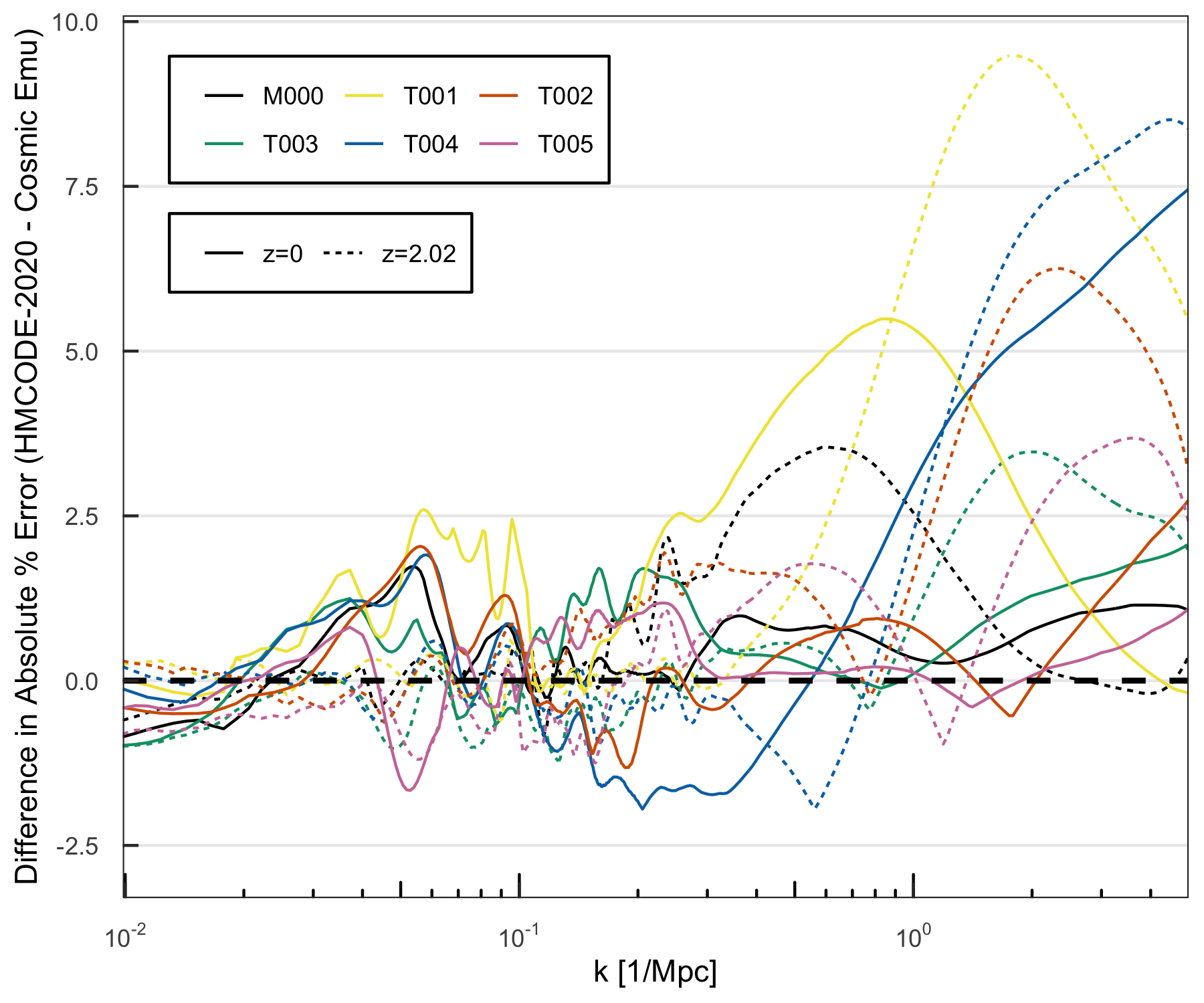}
\caption{Difference in absolute percent error between HMCODE-2020 and our emulator. Values above 0 indicate that HMCODE-2020 has relatively worse performance (i.e., higher absolute \% error) than our emulator.\label{fig:perf_comp_supp}}
\end{figure}

A direct comparison of the accuracy of the CosmicEmu and HMCODE-2020 for each cosmology and redshift combination is shown in Figure \ref{fig:perf_comp_supp}. Our emulator outperforms HMCODE-2020 in terms of absolute error integrated over logged $k$ for all test cosmologies and redshifts; that is, the integral of each curve in Figure~\ref{fig:perf_comp_supp} is greater than 0 for all assessed cosmologies at both redshifts, as calculated via the \texttt{pracma::trapz()} function in \textbf{R}.

\section{Cross Validation}

\begin{figure}
\centering
\includegraphics[width=0.45\textwidth]{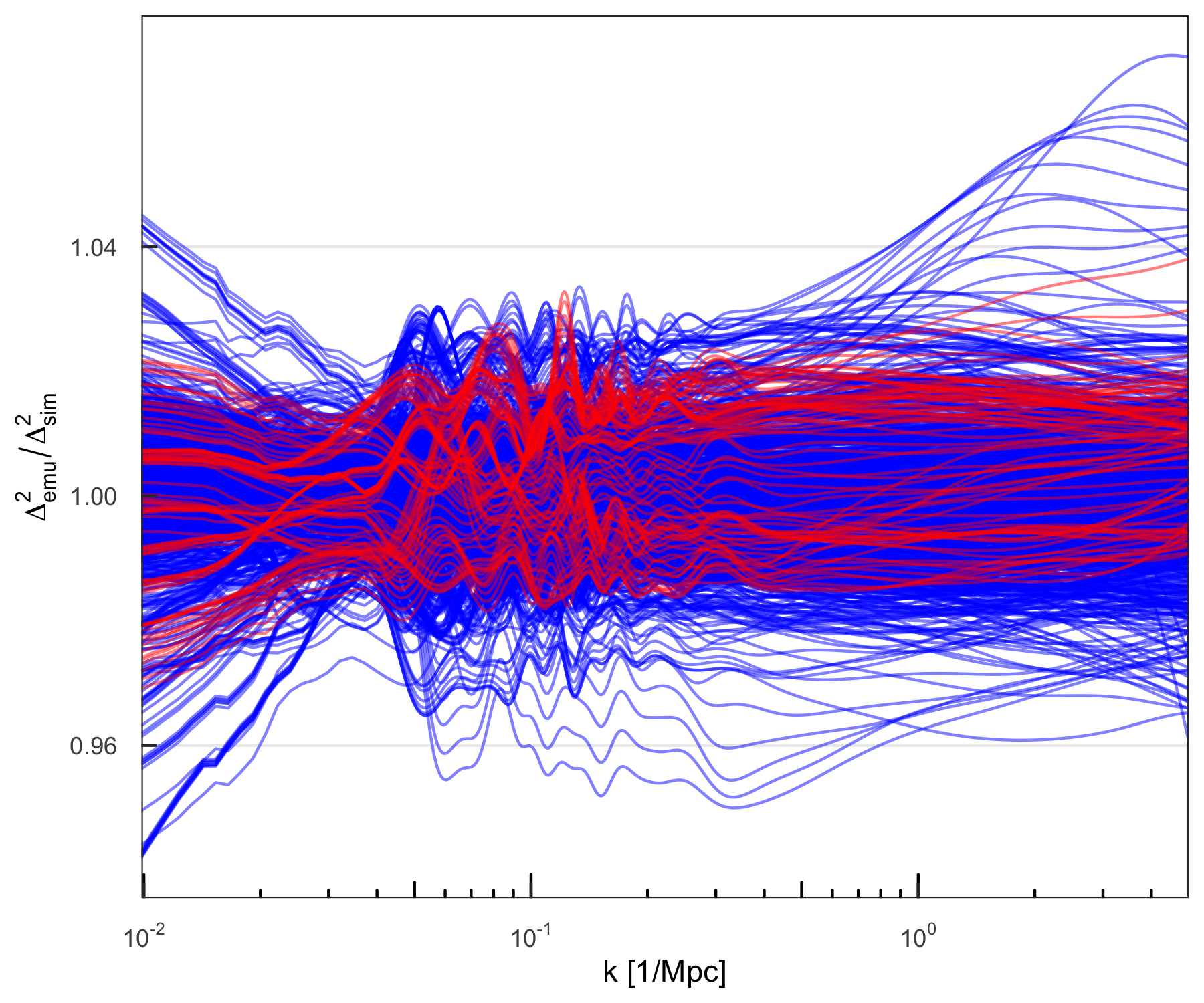}
\includegraphics[width=0.45\textwidth]{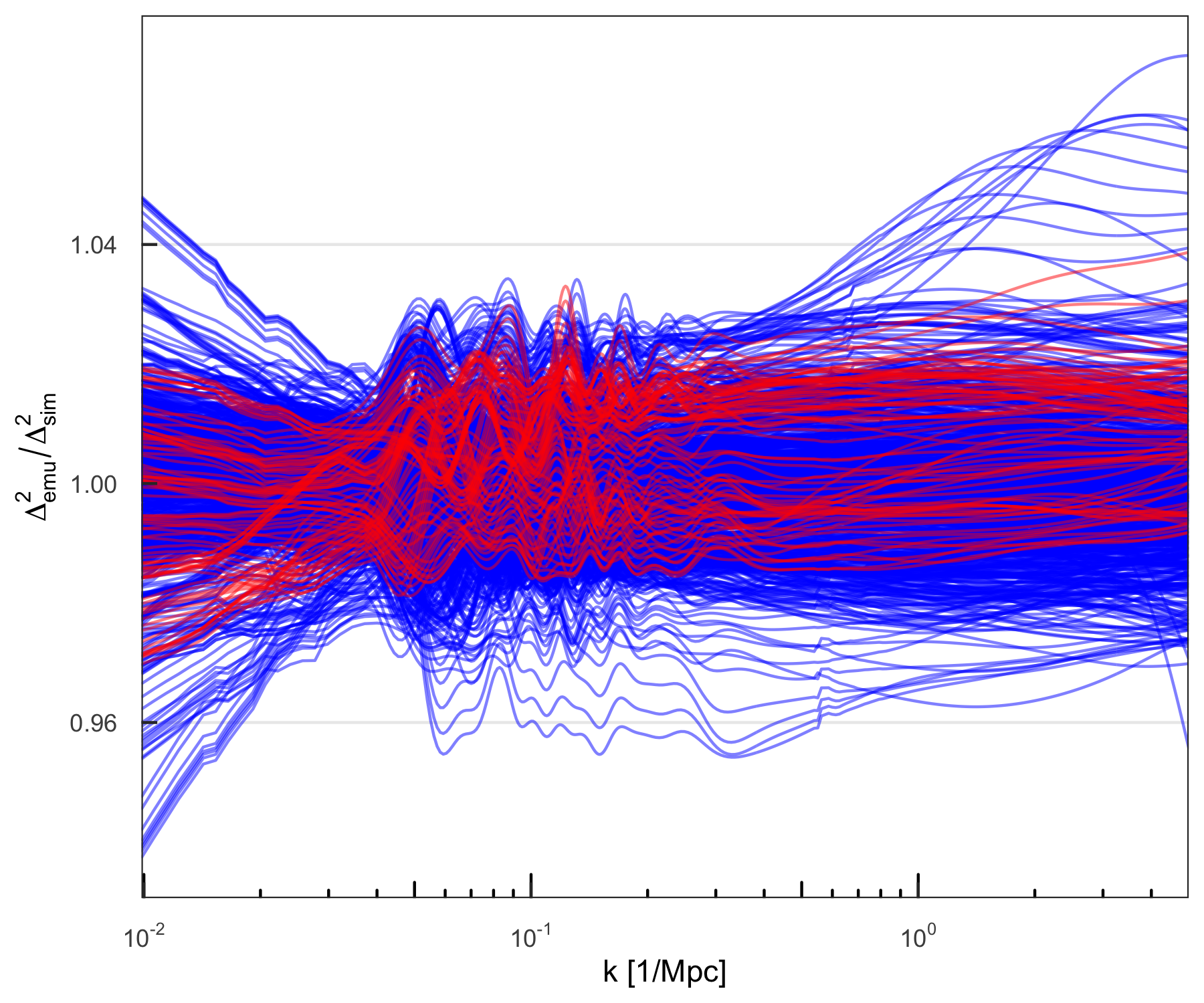}
\caption{Leave-out-one cross-validation results for the $P_{cb}$ emulator (left) and the $P_{tot}$ emulator (right) on all cosmologies. The ratio $\Delta^2_{\text{emu}} / \Delta^2_{\text{sim}}$ ranges from 0.941 to 1.071 for $P_{cb}$, and from 0.937 to 1.072 for $P_{tot}$. Red (blue) lines correspond to cosmologies having (non)zero neutrino mass.\label{fig:perf_crossval_all}}
\end{figure}

Figure \ref{fig:perf_crossval_all} shows leave-out-one cross-validation results for the $P_{cb}$ and the $P_{tot}$ emulators on all cosmologies included in the model. Accuracy is better than 4\% for most cosmologies (3\% for cosmologies having zero neutrino mass), with worst-case performance for cosmologies on the fringes of the design space.


\clearpage

\end{document}